\documentclass{jpp}
\usepackage{graphicx}

\usepackage[utf8]{inputenc}
\usepackage[T1]{fontenc}
\usepackage{amsmath}
\usepackage{xcolor}

\shorttitle{On low-n Mercier instabilities in Wendelstein stellarators}
\shortauthor{R. Ramasamy, H. Zhang, J. Geiger, et al.}

\title{A comparison of low-n Mercier unstable Wendelstein stellarators and quasi-interchange modes in tokamaks}

\author{Rohan Ramasamy\aff{1}\corresp{\email{rohan.ramasamy@ipp.mpg.de}}, Haowei Zhang\aff{1}, Joachim Geiger\aff{2}, Carolin N\"uhrenberg\aff{2}, H\r{a}kan M. Smith\aff{2}, Karl Lackner\aff{1}, Valentin Igochine\aff{1} and the JOREK team\aff{3}}

\affiliation{\aff{1} Max-Planck Institut f\"ur Plasmaphysik, Boltzmannstraße 2, 85748 Garching bei München, Germany
\aff{2} Max-Planck Institut f\"ur Plasmaphysik, Wendelsteinstrasse 1, 17491 Greifswald, Germany
\aff{3} see list of contributors in \cite{hoelzl2021jorek} }

\begin{document}

\maketitle

\begin{abstract}

Mercier's criterion is typically enforced as a hard operational limit in stellarator design. At the same time, past experimental and numerical studies have shown that this limit may often be surpassed, though the exact mechanism behind this nonlinear stability is not well understood. This work aims to contribute to our current understanding by comparing the nonlinear evolution of Mercier unstable Wendelstein stellarators to that of nonlinearly stable quasi-interchange modes in tokamaks. A high mirror, very low $\iota$, W7-X-like configuration is first simulated. Broad flow structures are observed, which produce a similar MHD dynamo term to that in hybrid tokamak discharges, leading to flux pumping. Unlike in tokamaks, there is no net toroidal current to counterbalance this dynamo, and it is unclear if it can be sustained to obtain a similar quasi-stationary nonlinear state. In the simulation, partial reconnection induced by the overlap of multiple interchange instabilities leads to a core temperature crash. A second case is then considered using experimental reconstructions of intermediate $\beta$ W7-AS discharges, where saturated low-n modes were observed experimentally, with sustained MHD signatures over tens of milliseconds. It is shown that these modes do not saturate in a benign quasi-stationary way in current simulations even in the presence of background equilibrium $\mathbf{E} \times \mathbf{B}$ flow shear. This leads to a burst of MHD behaviour, inconsistent with the sustained MHD signatures in the experiment. Nevertheless, the (1, 2) mode is observed at the experimental Spitzer resistivity, and its induced anomalous transport can be overcome using an experimentally relevant heat source, reproducing these aspects of the dynamics. The possible reasons for the discrepancies between experiment and simulation, and the observation of partial reconnection in contrast to flux pumping are discussed, in view of reproducing and designing for operation of stellarators beyond the Mercier stability limit.
\end{abstract}

\section{Introduction} \label{sec:intro}


In recent stellarator optimisation efforts, the design requirements for improved turbulent transport properties and quasi-symmetry \citep{roberg2024reduction, landreman2022mapping} have been found to conflict with the need for magnetohydrodynamic (MHD) stability, as defined by Mercier's criterion \citep{mercier1964equilibrium}. For quasi-isodynamic stellarators \citep{goodman2024quasi}, while optimisation for the maximum-\( \mathcal{J} \) property leads to good magnetohydrodynamic stability properties, it is also known that the Mercier stability constraint often prevents access to attractive configurations, particularly with respect to coil complexity. 

The conflict of requirements in the stellarator optimisation community has led many to ask whether it is possible to remove or relax the constraint presented by Mercier's criterion. Hope that this may be possible has been informed by several experiments, which have successfully operated beyond the $\beta$ limit prescribed by Mercier stability \citep{de2015magnetic, weller2003experiments, weller2006significance, ohdachi2017observation}. The experimental analysis has been supported by several encouraging nonlinear MHD and gyrokinetic studies of existing and conceptual scenarios \citep{mishchenko2023global, sato2021kinetic, zhou2024benign}. Particularly, in the case of ballooning modes, two fluid effects have long been anticipated to stabilise and mitigate the induced transport from such MHD activity \citep{strauss2004simulation}.  At the same time, there are also examples of Mercier unstable configurations which do indeed suffer from \textcolor{black}{degraded confinement due to MHD activity driven by interchange modes \citep{wright2024investigating, sakakibara2010study}}. As a result, further experimental and simulation work is warranted to understand when and to what extent Mercier's criterion is necessary.

In contrast to stellarators, Mercier's criterion is rarely discussed in tokamak design. This is because the hard limit prescribed by current driven kink modes is normally more restrictive than that prescribed by Mercier's criterion \citep{zohm2014magnetohydrodynamic}, such that ballooning and combined current and pressure driven modes are the main cause for concern. An exception to this is in the core of the device for hybrid scenarios, where interchange modes have been proposed as a potential candidate for sawtooth crashes \citep{wesson1986sawtooth}. Importantly, nonlinearly stable MHD states have been observed in hybrid tokamak scenarios which can prevent sawtooth cycles via the onset of quasi-interchange modes, and lead to improved performance \citep{burckhart2023experimental}. Very recently, such states have been shown even in DIII-D with strong negative triangularity - an operational scenario typically hampered by relatively violent MHD activity - leading to sustained high performance \citep{boyes2024novel}. 

The typical mechanism to explain this behaviour is flux pumping \citep{petty2009magnetic}. In the case of the quasi-interchange mode, the flux pumping mechanism is attributed to a MHD dynamo term, which acts to redistribute the ohmically driven current in tokamaks, keeping the q profile close to unity \citep{jardin2020new, krebs2017magnetic}. \textcolor{black}{For this kind of scenario, current control is required to maintain moderate externally driven current density peaking, which the MHD dynamo term is able to redistribute.} In such a way, the current density gradients driving the internal kink mode can be flattened in the core, such that sawteeth and crash dynamics are prevented. 

\begin{figure}
\includegraphics[width=\linewidth]{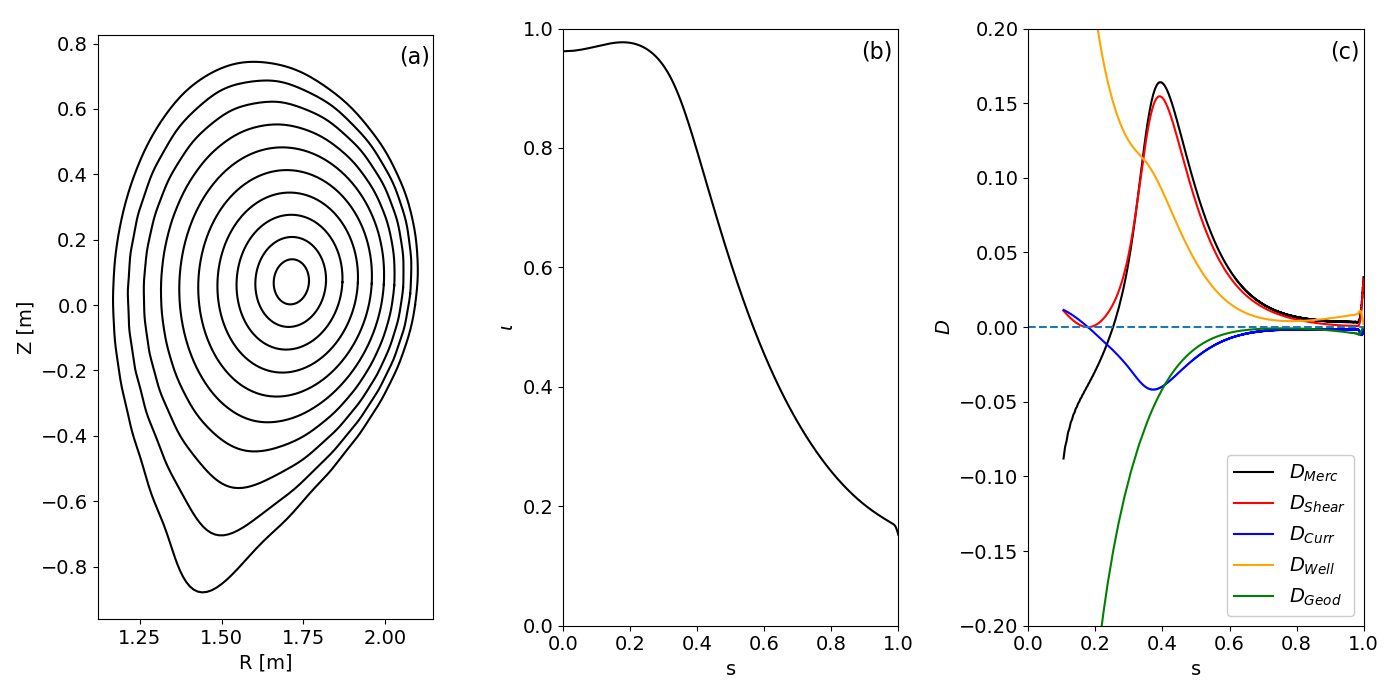}
\caption{Equilibrium flux surfaces (a) and $\iota$ profile (b) of an ASDEX upgrade discharge characteristic of flux pumping, reconstructed using VMEC. The contributions to Mercier's criterion (c) computed in VMEC imply that the plasma is Mercier unstable in the core of the device. ($D_{Merc} < 0$). The VMEC computation was carried out with 1001 radial points.}
\label{fig:aug_mercier_stability}
\end{figure}

In order to demonstrate the link between flux pumping and Mercier stability explicitly, a VMEC  \citep{HirshmanRij1986} reconstruction of an ASDEX Upgrade discharge characteristic of flux pumping scenarios has been assessed in terms of its Mercier stability in Figure \ref{fig:aug_mercier_stability}. This equilibrium has also been simulated nonlinearly, reproducing the flux pumping phenomenon with realistic equilibrium and plasma parameters for the first time \citep{zhang2024full}. In this and all subsequent figures, $s=\sqrt{\psi_t}$, corresponds to the square root of the \textcolor{black}{normalised} toroidal flux. It can be seen that this case is a nonlinearly Mercier stable mode, according to experiment and simulation.

Past studies have shown that such helical states are not unique to tokamaks, and occur also in Reverse Field Pinches \citep{piovesan2017role}. A natural question is therefore to ask whether such a regime can be obtained in a stellarator? This contribution aims to provide an initial understanding of this via nonlinear MHD analysis of Wendelstein stellarators using the JOREK code \citep{huysmans2009non, hoelzl2021jorek}. The paper is outlined as follows. Section \ref{sec:w7x} characterises the simulation of a low n interchange mode in a W7-X-like plasma, demonstrating an appreciable dynamo term is observed that could in principle lead to a flux pumping scenario. The nonlinearly simulated mode is shown to partially reconnect, however, leading to core temperature crash dynamics. 

Turning to an experimental observation of a saturated low n instability, Section \ref{sec:w7as} aims to observe whether low n saturated MHD modes in W7-AS can be re-produced by simulations. It is shown that in current simulations, while low n modes are observed, in the absence of $\mathbf{E} \times \mathbf{B}$ flows, multiple competing transient modes are present. This is inconsistent with the coherent (n=1, m=2) mode found in the experiment. These interacting modes drive a loss of the stored plasma thermal energy. Including background flows, high n modes above n > 5 are suppressed by \textcolor{black}{equilibrium} shear flow stabilisation, leading to a dominant (5, 10) mode structure. A sustained (1, 2) mode structure has still not been observed in simulations, though such a mode can be observed transiently depending on the initialisation of the simulation. The stored energy can be sustained within the experimental heating power, for the prescribed diffusive transport coefficients which are intended to approximate the experiment. 

Section \ref{sec:discuss} interprets and outlines what can be concluded from the above results. \textcolor{black}{In particular, it is argued that the reason a clear flux pumping scenario is not currently observed is because two key features of such discharges are missing in the current simulation set up, namely that (1) a modest amount of current drive and control is used, and (2) a single marginal instability is present in the mode spectrum.} Section \ref{sec:conclusion} summarises the study and suggests paths for further development.

\section{Assessment of interchange modes in a W7-X-like configuration} \label{sec:w7x}

The first case studied numerically is a very low $\iota$, high mirror W7-X-like configuration. This configuration was originally studied as part of the candidate configurational space of W7-X \citep{nuhrenberg1996global}. The final configurational space of W7-X was later constrained so that the simulated central $\iota$ value lies outside of its operational range. \textcolor{black}{In such a way, the analysis in this Section is not intended to be compared directly with results from the W7-X experiment.} Nevertheless, the case has many similarities to W7-X plasmas with high mirror, and low $\iota$, such as the TEH configuration studied in \citep{zhou2024benign}. The configuration was initially chosen for simulation as a test case of a violent mode representing a worst case for the dynamics in W7-X.

\subsection{Equilibrium and linear stability}
\begin{figure}
\includegraphics[width=0.495\linewidth]{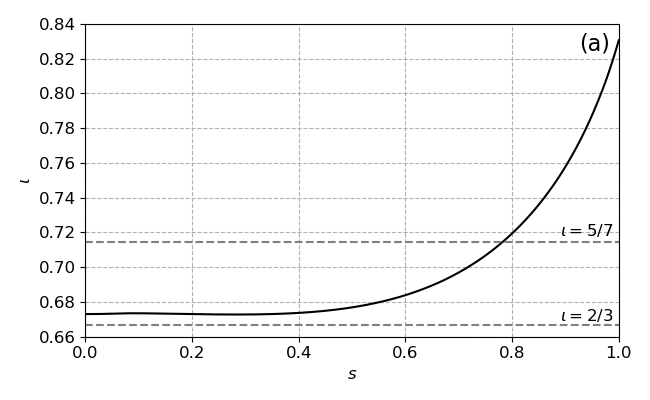}
\includegraphics[width=0.495\linewidth]{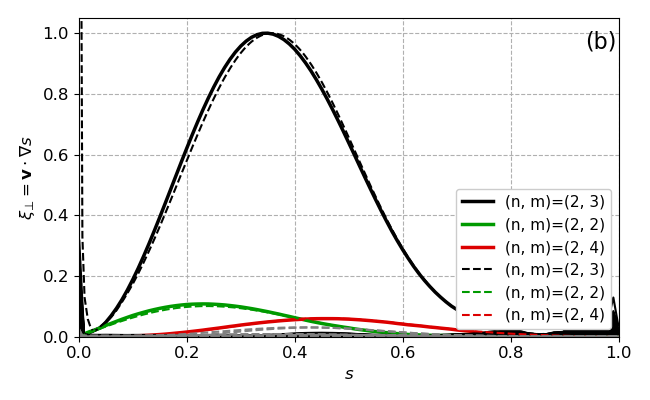}
\includegraphics[width=0.66\linewidth]{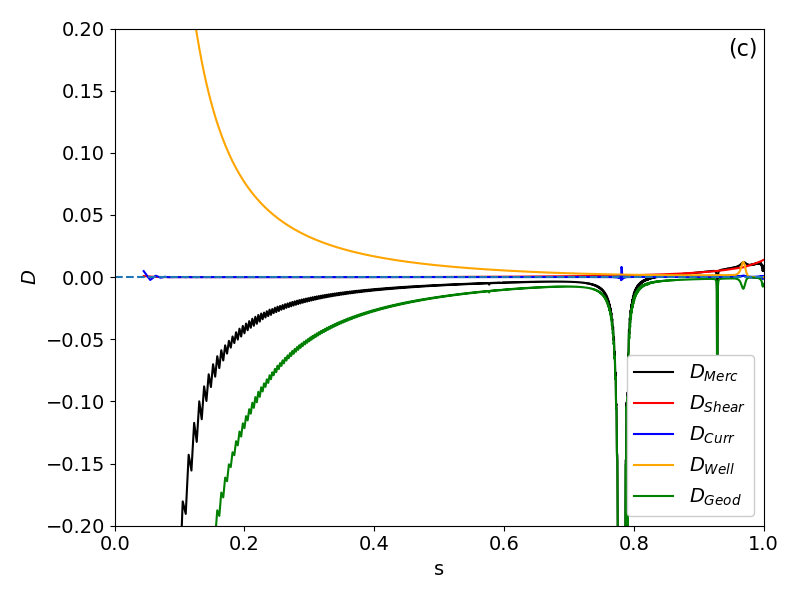}
\includegraphics[width=0.33\linewidth]{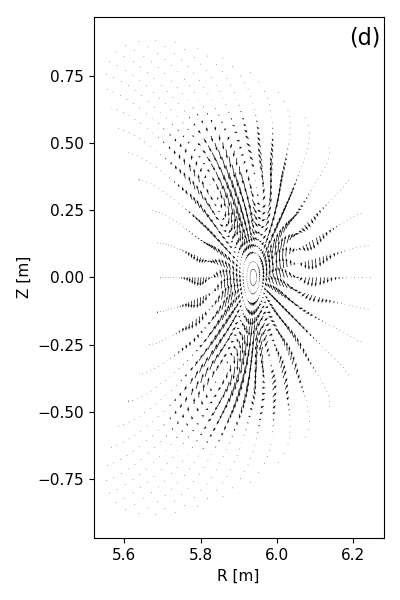}
\caption{Equilibrium and linear stability properties of a very low $\iota$, high mirror W7-X-like configuration. The $\iota$ profile (a) shows a 2/3 resonance is close to the $\iota$ value on the magnetic axis. The equilibrium is linearly MHD unstable to a (2, 3) interchange mode in both JOREK and CASTOR3D (b). Mercier's criterion shows that the configuration is interchange unstable over most of the plasma volume (c). The velocity vector plot (d) shows similar broad structures within the instability region as observed in Figure 14 of \citep{krebs2017magnetic}.}
\label{fig:w7x_linear}
\end{figure}

The $\iota$ profile and ideal MHD stability properties of this case are shown in Figure \ref{fig:w7x_linear}. It can be seen that the central $\iota$ profile is flat and lies very close to the $2/3$ rational surface. This leads to broad unstable modes with this helicity, in particular the (2, 3) interchange. The case is shown to be Mercier unstable, according to the diagnostic computed in VMEC. The structure of the separate contributions to Mercier's criterion in Figure \ref{fig:w7x_linear} have a similar form to other low shear magnetic hill configurations, such as those studied in \citep{de2015magnetic}. The plasma is unstable over a much larger region of the enclosed volume than in the AUG flux pumping case in Figure \ref{fig:aug_mercier_stability}. 

Linear MHD analysis in both JOREK and CASTOR3D demonstrate that the configuration is unstable to both low and high $n$ \textcolor{black}{(not shown)} ideal interchange modes, similar to the original results in \citep{nuhrenberg1996global}. \textcolor{black}{It is important to note that, similar to the observations in \citep{ramasamy2024nonlinear}, the correct treatment of helical mode coupling and fluid compressibility in the reduced MHD model in JOREK, outlined in Appendix \ref{app:reduced_mhd}, is necessary in order to observe these modes. These compressible effects are included in all of the results shown in this paper.}

In nonlinear MHD simulations of stellarators, it can be difficult to observe saturated low n pressure driven modes because their high n counterparts often saturate first in the absence of stabilisation mechanisms such as sheared flow rotation, significant parallel heat transport\textcolor{black}{, or finite larmor radius effects}. This is partly because of the larger ideal MHD growth rate of high n modes. In addition however for stellarator simulations, the finite initial equilibrium force balance error introduced from the numerical import from ideal MHD solvers like GVEC \citep{hindenlang2017parallel} means that the kinetic energy of $N_f=0$ modes typically reaches a moderate kinetic energy within a few Alfven times. The initial energy is not enough to drive macroscopic dynamics, but means that the modes have a head start towards nonlinear saturation. For these reasons, in the simulated W7-X-like case, the (10, 15) and other $N_f=0$ modes are typically observed to saturate before the (2, 3) mode. To circumvent this issue in the results presented herein, the $N_f=2$ mode family is allowed to evolve and grow without updating the $N_f=0$ and $N_f=1$ mode families in the early stage of the simulation, $t < 0.85\ ms$\textcolor{black}{, where the equilibrium profiles are approximately preserved}. As the $N_f=2$ kinetic energy approaches but remains broadly decoupled from the background $N_f=0$ modes, the other mode families are then updated in time to simulate the full spectrum as the nonlinear phase begins. This allows one to observe the nonlinear dynamics in the case where the low $n$ modes lead the dynamics, which is the case of interest for this study. 

The $n=2$ velocity perturbation associated with the linearly dominant (2, 3) mode observed with JOREK is compared with the corresponding instability computed in the linear viscoresistive full MHD code CASTOR3D \citep{Strumberger2016} in Figure \ref{fig:w7x_linear}. Good agreement is observed in the linear mode structure. Reconstructing the linear flow pattern in CASTOR3D, it can be seen that the mode has the characteristic broad poloidal flow structures associated with flux pumping.

\subsection{Nonlinear dynamics and partial reconnection}
The simulation parameters used in the nonlinear simulations presented in this Section are outlined in Appendix \ref{app:params}. This simulation is not intended to reproduce an experimental discharge. The relatively high viscosity is chosen in order to improve the numerical stability of the case, which is challenging to simulate in the no viscosity limit. Similar to experience from the FAR3D code \citep{varela2024stability}, numerical stability tends to improve as the number of toroidal harmonics included in simulations is increased. Full torus simulations with more than 30 toroidal harmonics are currently prohibitively expensive for doing parameter scans, and so $n_{tor}=30$ was chosen as a reasonable compromise. Figure \ref{fig:w7x_linear} (b) shows that the grid resolution parameters do not have a significant influence on the mode structure of the (2, 3) mode.

\begin{figure}
\centering
\includegraphics[width=0.95\linewidth]{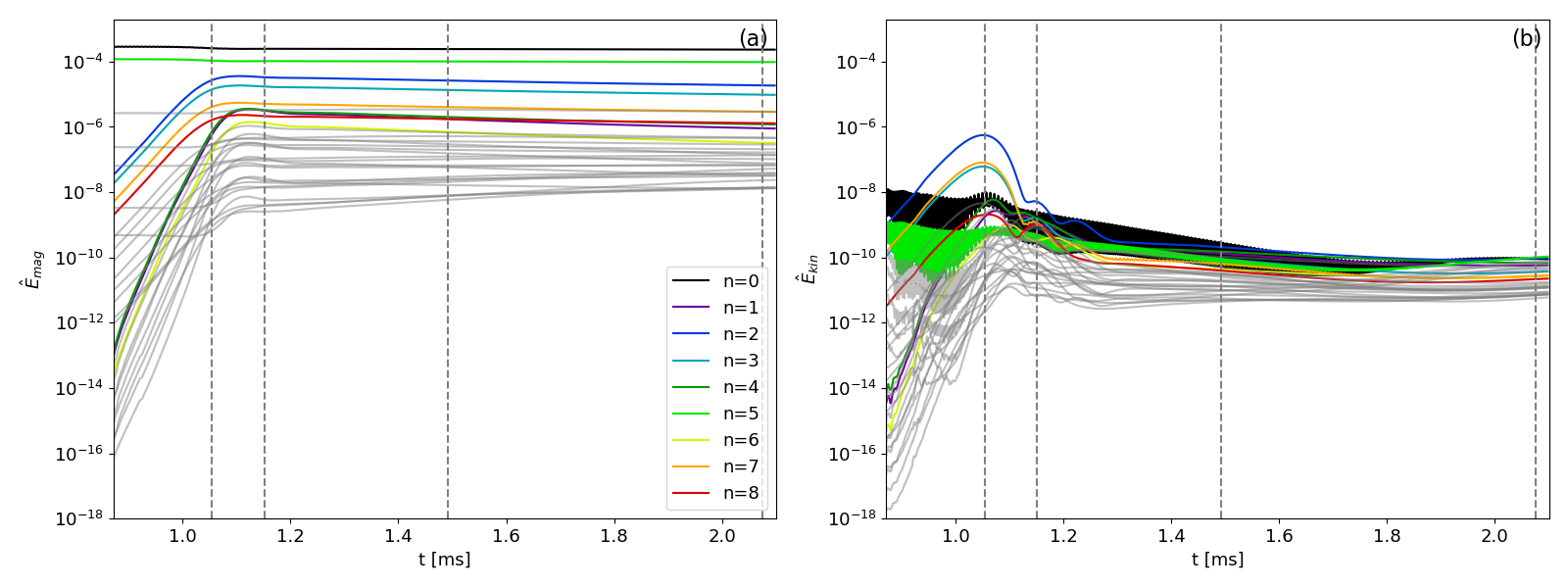}
\includegraphics[width=0.95\linewidth]{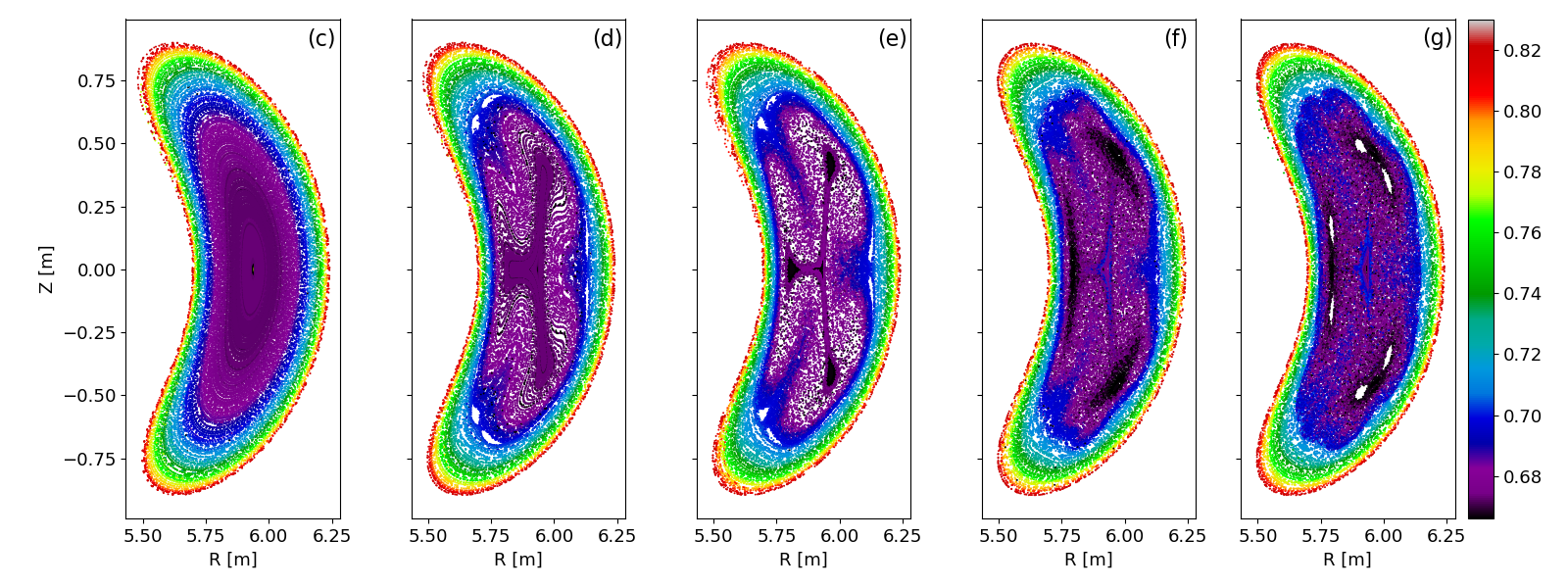}
\includegraphics[width=0.17\linewidth]{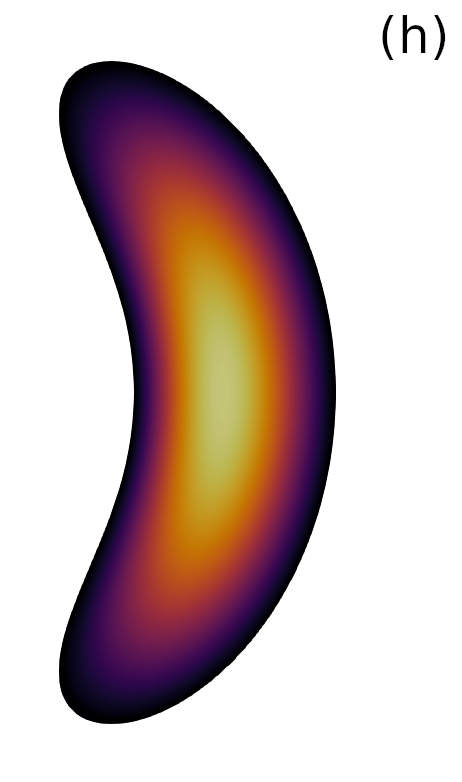}
\includegraphics[width=0.17\linewidth]{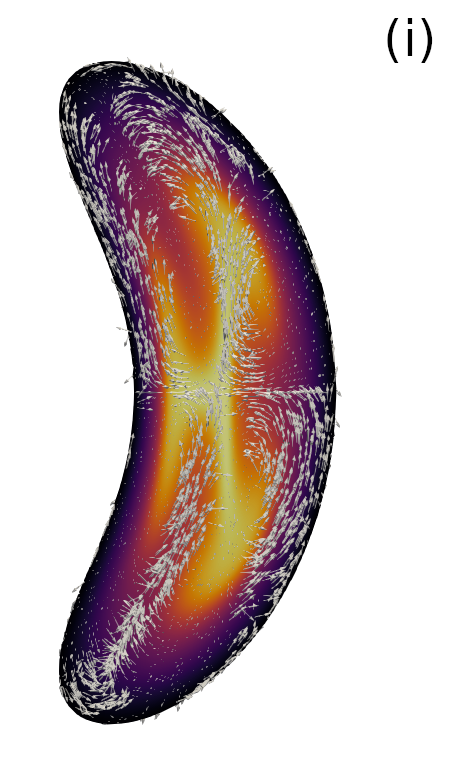}
\includegraphics[width=0.17\linewidth]{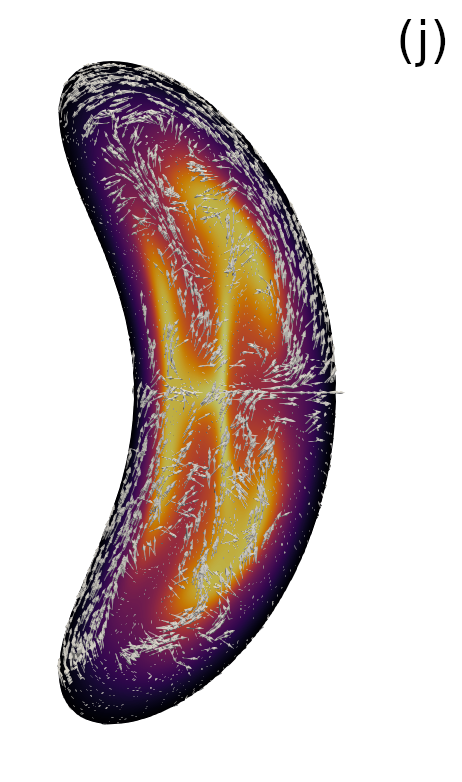}
\includegraphics[width=0.17\linewidth]{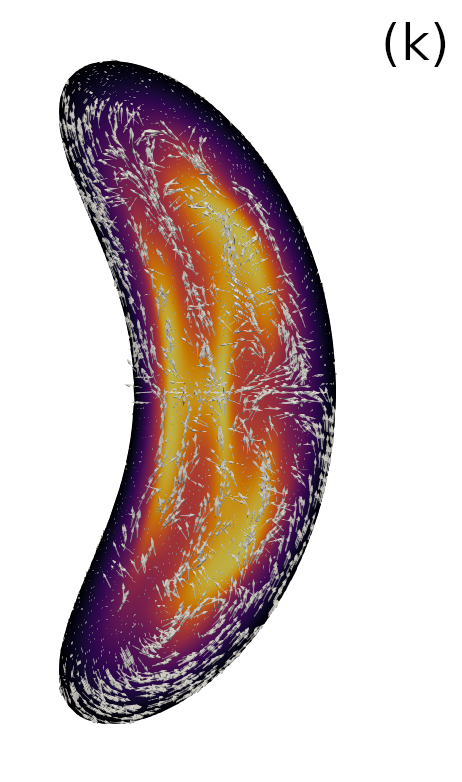}
\includegraphics[width=0.17\linewidth]{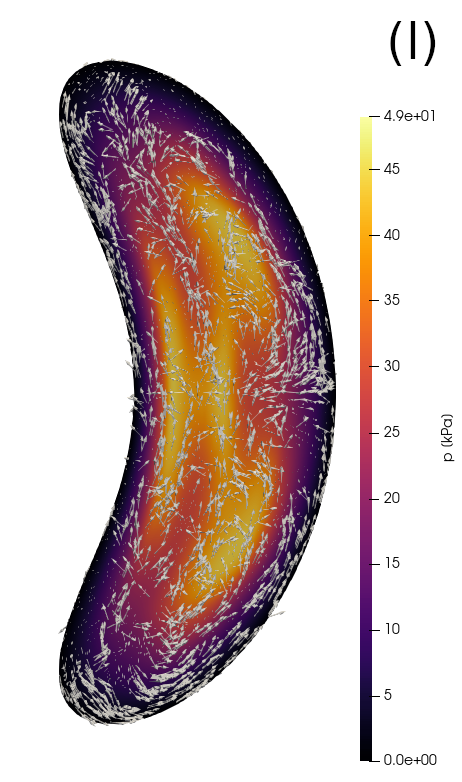}
\caption{Evolution of magnetic (a) and kinetic (b) energies during the initial nonlinear phase of a (2, 3) interchange mode in a W7-X-like configuration. The initial dynamics of the dominant (2, 3) perturbation is ideal (c-d), as illustrated by the nested compression of flux surfaces. Only in the late nonlinear phase (e-g), the local $\iota$ value changes due to reconnection of local current sheets. \textcolor{black}{The pressure profile deformation (h-l) is approximately aligned with the flux surface deformation.}}
\label{fig:w7x_nonlinear}
\end{figure}

\begin{figure}
    \centering
    \includegraphics[width=0.495\linewidth]{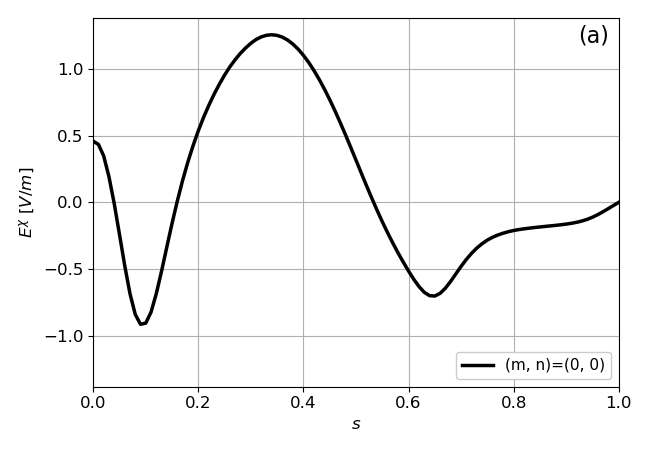}
    \includegraphics[width=0.495\linewidth]{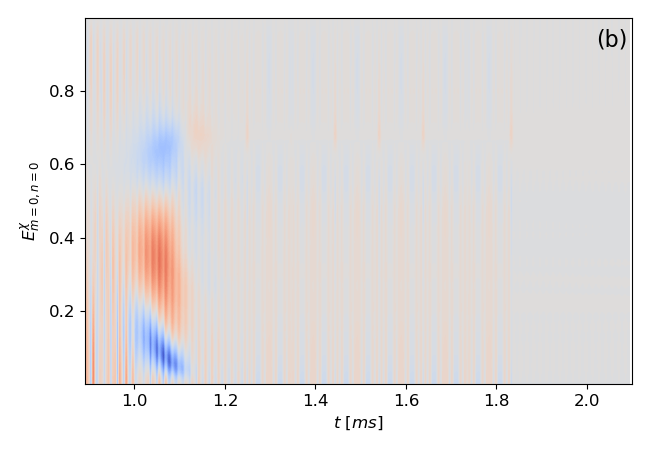}
    \caption{Radial profile of the (0, 0) component of the electric field (a) induced along $\nabla \chi$\textcolor{black}{, the vacuum magnetic field as described in Appendix \ref{app:reduced_mhd},} due to the dynamo voltage. This profile is plotted over time (b) to show that the dynamo is not sustained on the resistive timescale.}
    \label{fig:w7x_E_chi}
\end{figure}

Dynamics of the nonlinearly evolved mode in JOREK are shown in Figure \ref{fig:w7x_nonlinear} and \ref{fig:w7x_E_chi}. In the simulation shown, the (2, 3) interchange saturates first. The Poincare plots in Figure \ref{fig:w7x_nonlinear} are coloured by the corresponding $\iota$ value, demonstrating that the initial mode is ideal, because the $\iota$ value of the core flux surfaces remains unchanged during the early saturation of the mode in Figure \ref{fig:w7x_nonlinear} (c-d). The corresponding electric field due to the $\mathbf{v} \times \mathbf{B}$ dynamo is computed in Figure \ref{fig:w7x_E_chi}. The vacuum magnetic field direction is used for this projection of the electric field, as it is assumed to be a suitable approximation of the total magnetic field. It can be seen that a dynamo term is produced by the flow structures which has a similar $m=0,\ n=0$ component to the (1, 1) quasi-interchange in tokamaks \citep{krebs2017magnetic, piovesan2017role}. Plotting the evolution of the dynamo voltage in time, the electric field is not sustained in the late nonlinear phase $t > 1.15\ ms$. A faint oscillation is observed in the dynamo, which is a weaker variant of what has been observed in tokamak simulations, and correlates with the weak oscillation observed in the $n=2$ kinetic energy. The faster dissipation of this oscillation could be because of the stronger spatial gradients associated with a (2, 3) mode, compared to the typical (1, 1) in tokamaks. In any case, the oscillation and dissipation of the initial fast dynamo has also been observed in tokamak simulations \citep{zhang2024full}, giving way to a resistive slow dynamo that self organises on the longer resistive time scale and is sustained for 100s of milliseconds.

\begin{figure}
    \centering
    \includegraphics[width=0.495\linewidth]{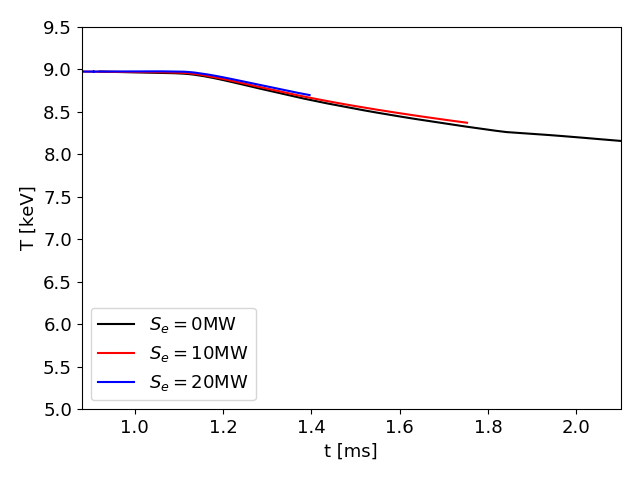}
    \caption{Time trace of the core temperature for simulations at different heating power show a crash of the temperature profile.}
    \label{fig:w7x_temp_crash}
\end{figure}

Observations of a resistive slow dynamo could not be reproduced in the current test case, because of strong reconnection in the late nonlinear phase leading to crash dynamics. The violent dynamics of the (2, 3) mode lead to significant compression of the flux surfaces, implying the formation of current sheets. The reconnection observed in the third to fifth panels from $t > 1.15\ ms$ is a secondary effect of the ideal mode, as the current sheet breaks apart the original nested flux surfaces into reconnection sites, where the $\iota$ profile shows the local $\iota$ is modified to follow the $2/3$ rational surface. Later in time, after $t > 1.2\ ms$, the region around the dominant (2, 3) mode starts to become stochastic, as other modes grow and compete with the dominant mode structure. 

\textcolor{black}{Regarding the thermal energy transport, the pressure plots in Figure \ref{fig:w7x_nonlinear} (h-l) show that the pressure contours remain approximately aligned with the deformed flux surfaces, implying that diffusive transport is sub-dominant. This can also be seen from the overlaid velocity vectors in these plots, which are correlated with the outward motion of the plasma core, and vortical structures required for the dynamo observed in Figure \ref{fig:w7x_E_chi}. It should be noted that these velocity vectors are normalised at the plotted time step for better visibility, and so cannot be quantitatively compared between plots.} Considering the time trace for the core temperature in Figure \ref{fig:w7x_temp_crash}, the temperature drops significantly on a time scale of $1\ ms$. A uniform heat source comparable to the experimental heating power in W7-X is not able to prevent the sharp drop in temperature, making it comparable to a crash.

It is important to observe that unlike in sawteeth, the $\iota$ value in the core of the device does not transition to $2/3$ across the plasma volume. This implies that the reconnection process is partial. Partial reconnection is associated with a mechanism for high $\beta$ disruptions in tokamaks, where it is argued that the overlap of multiple interchange modes of different toroidal mode number can lead to a sawtooth crash \citep{itoh1995sawtooth, jardin2020new}. While the multiple heat exhaust points from the core that are observed for such crashes \citep{liu2024discriminating} conflict with the typical experimental observation in tokamaks of a local exhaust point, the model is a valid mechanism for crash dynamics. Particularly in the case of stellarator configurations which are unstable to multiple interchange modes, partial reconnection could drive a crash of the core temperature.

The partial reconnection and significant drop in core temperature prevent the interrogation of the dynamo loop voltage and flux pumping mechanism over long time scales, as is typically attempted in tokamak studies of flux pumping, where the dynamo signature is sustained over 100s of milliseconds - significantly longer than the simulated time of this stellarator case. In such a way, the observation of partial reconnection poses an interesting challenge to the possibility of achieving flux pumping in a stellarator. In tokamaks, flux pumping requires the plasma flows to self-organise into a coherent quasi-stationary dynamo that counteracts modifications to the plasma state induced by current sources - if the current source strengthens or weakens, the linear growth rate of the mode is modified, and the dynamo strengthens or weakens in turn to reverse the effect of the current source. In such a way, a quasi-stationary state can be sustained throughout the nonlinear phase, on the timescale of hundreds of milliseconds. This process only works within a certain window of operation. If the current source is too strong, the profiles will be modified to reduce the linear growth rate of the interchange. The dynamo will then not be able to compete. In the case of a tokamak, a sawtooth will then occur as the current density continues to rise, crossing $\iota=1$. 

Reconnection and stochasticity is a mechanism for decorrelating the dominant dynamo structure, due to redistribution from parallel transport. Any reconnection that occurs must therefore remain a subdominant process if the mode is to be sustained over resistive time scales. This is more likely to occur for modes closer to marginality than the simulated case, which is strongly unstable. As suggested in \citep{jardin2020new}, a crash driven by partial reconnection may also be prevented if the case is stable to all but a single MHD mode, such that coherent MHD structures and a dynamo can form on the resistive time scale. It should also be noted that ASDEX Upgrade studies of flux pumping at experimentally relevant parameters also observe that only the (1, 1) mode is linearly unstable in flux pumping discharges \citep{zhang2024full}. Note that for the stellarator case, a single mode does not refer to single m and n quantum numbers, but instead includes the helical coupling of different poloidal and toroidal modes. These modes are combined into a single MHD instability, comprised of multiple harmonics \citep{schwab1993ideal}. 

\section{W7-AS low $n$ pressure driven modes at intermediate $\beta$} \label{sec:w7as}

In this Section, an alternative perspective on low n Mercier unstable Wendelstein configurations is pursued by attempting to reproduce the low n MHD activity observed in experimental W7-AS discharges at intermediate $\beta$ values \citep{weller2003investigation}. In such discharges, a clear (1, 2) mode structure has been observed in X-ray tomograms, which rotates at a frequency that roughly corresponds to the background $\mathbf{E} \times \mathbf{B}$ flow. At these $\beta$ values, the plasma is known to be Mercier and resistive interchange unstable. As $\beta$ is increased, the (1, 2) mode activity can be suppressed, which could indicate that flux pumping plays a role in the dynamics. Flux pumping can be seen not only as a final, but also as a transient state, which provides stability on the way to a new configuration, which might be more difficult, or impossible to reach without traversing the nonlinearly stable state, facilitated by this nonlinear stabilisation mechanism.

The current work builds on the author's previous attempts in \citep{ramasamy2024nonlinear}, where a series of validation studies were conducted against past experimental observations from W7-AS. This past validation effort was only partially successful, as a saturated (1, 2) mode was not clearly observed with approximate experimental parameters.

\subsection{Equilibrium and simulation set up with background $\mathbf{E} \times \mathbf{B}$ flows} \label{sec:w7as_setup}
\begin{figure}
    \centering
    \includegraphics[width=0.32\linewidth]{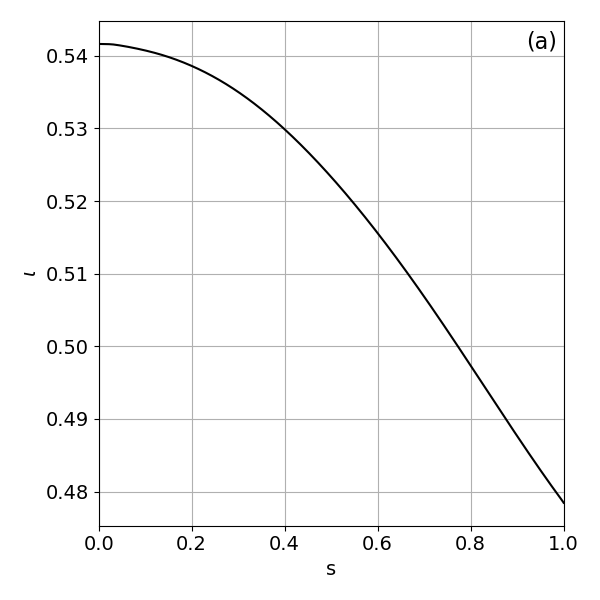}
    \includegraphics[width=0.32\linewidth]{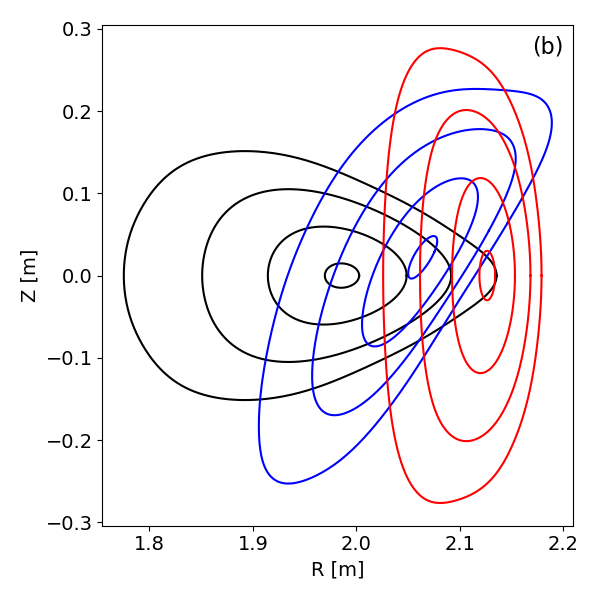}
    \includegraphics[width=0.32\linewidth]{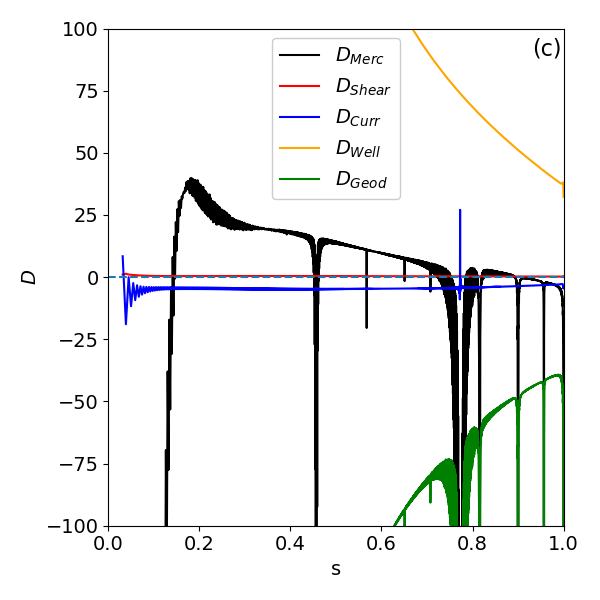}
    \caption{W7-AS experimental equilibrium reconstruction with $8\ kPa$ core pressure. The $\iota$ profile (a) shows the presence of a low order 1/2 resonance. Flux surfaces in the $\phi=0.0,\ \frac{\pi}{10}$ and $\frac{\pi}{5}$ planes (b) show the basic magnetic topology of the five field period stellarator. The Mercier stability (c) shows local interchange instability as the $\iota$ profile crosses different rational surfaces.}
    \label{fig:w7as_equil}
\end{figure}

The $8\ kPa$ W7-AS case is targeted from \citep{ramasamy2024nonlinear}, as it was the simulated $\beta$ value with the strongest MHD activity. The equilibrium and its Mercier stability are shown in Figure \ref{fig:w7as_equil}. The plasma has a relatively low but finite magnetic shear, and a $\iota=0.5$ surface in the outer mid-radius, which is expected to be the dominant resonance. Compared to the W7-X-like case in Section \ref{sec:w7x}, the Mercier criterion is more locally unstable. It should be noted that this case is also resistive interchange unstable, and the experimental diagnostics suggest ballooning-like features \citep{weller2003experiments}.

Among the possible candidates for the past discrepancy between simulation and experiment in \citep{ramasamy2024nonlinear}, uncertainties in the resistivity and background ambipolar radial electric field were considered the prime candidates. The neoclassical resistivity was therefore computed and shown to not make a considerable difference when compared to the Spitzer resistivity. This is likely due to the relatively high collisionality of this cold, high density plasma, with $T \approx 180\ \mathrm{eV}$ and $n \approx 2 \times 10^{20}\ \mathrm{m^{-3}}$ in the core plasma\textcolor{black}{, corresponding to $\nu^* = O(1)$ across the plasma volume.} Further, it is known that the impurity content in the experimental discharges was low, such that the assumption that $Z_{eff} \approx 1.0$ is reasonable \citep{geiger2003analysis}. In such a way, the classical Spitzer resistivity is justified for use in simulations.

On the other hand, the experiment is known to have a significant ambipolar radial electric field, driving an equilibrium poloidal rotation profile. In the experiment, low frequency Mirnov coil signals indicate that the low $n$ mode activity rotates with this background flow. JOREK has recently been used to demonstrate the well established effect that this background flow should have a preferentially stabilising effect on high mode numbers in classical stellarators \citep{mcgibbon2024simulations}. \textcolor{black}{Introducing the equilibrium poloidal flow can therefore be expected to suppress the high n mode activity which was observed in the previous analysis in \citep{ramasamy2024nonlinear}. This can be expected to lead to improved confinement in a similar manner as diamagnetic flows can lead to Edge Harmonic Oscillation regimes in tokamaks, by suppression of high n ballooning modes \citep{chen2016rotational, feng2020optimization, brunetti2019excitation}.} In such a way, it could influence the simulation results for W7-AS, helping to reproduce the (1, 2) mode. 

A background flow can be initialised in JOREK by adding a poloidal momentum source to the system of equations, shown in Appendix \ref{app:reduced_mhd}. The momentum source maintains a background $\mathbf{E} \times \mathbf{B}$ field that is aligned to the initial equilibrium flux surfaces, allowing parallel Pfirsch-Schl\"uter flows to form self-consistently to preserve \textcolor{black}{$\nabla \cdot \mathbf{\rho v} = 0$ as the equations are evolved in time. It should be noted that, if the formation of these parallel flows is neglected, the compressibility of the fluid is not appropriately accounted for, by removing the parallel momentum equation from the system of equations in Appendix \ref{app:reduced_mhd}, artificial acoustic waves have been observed to pollute the solution. These acoustic modes lead to an artificial compression, which feeds back negatively on the applied poloidal momentum source, such that the prescribed poloidal flow field cannot be obtained. In such a way, it is necessary to appropriately account for the parallel Pfirsch-Schl\"uter flows in order to carry out these simulations.} 

The ambipolar radial electric field can be calculated from the equilibrium reconstruction using the assumed equilibrium profiles described in Appendix \ref{app:params}, and the DKES \citep{van1989variational} and Neotransp codes. As shown in \citep{Baldzuhn_1998}, the measured and calculated radial electric field can be inconsistent for $s \le 0.7$, however the authors did not find references for the pure NBI, high density discharges simulated herein. \textcolor{black}{It should be noted that for the intermediate collisionality of these discharges, turbulent and collisional effects may lead to deviations of the radial electric field in the plasma edge from the neoclassical limit assumed in the applied codes, which leads to higher modeling uncertainties.}

\begin{figure}
    \centering
    \includegraphics[width=\linewidth]{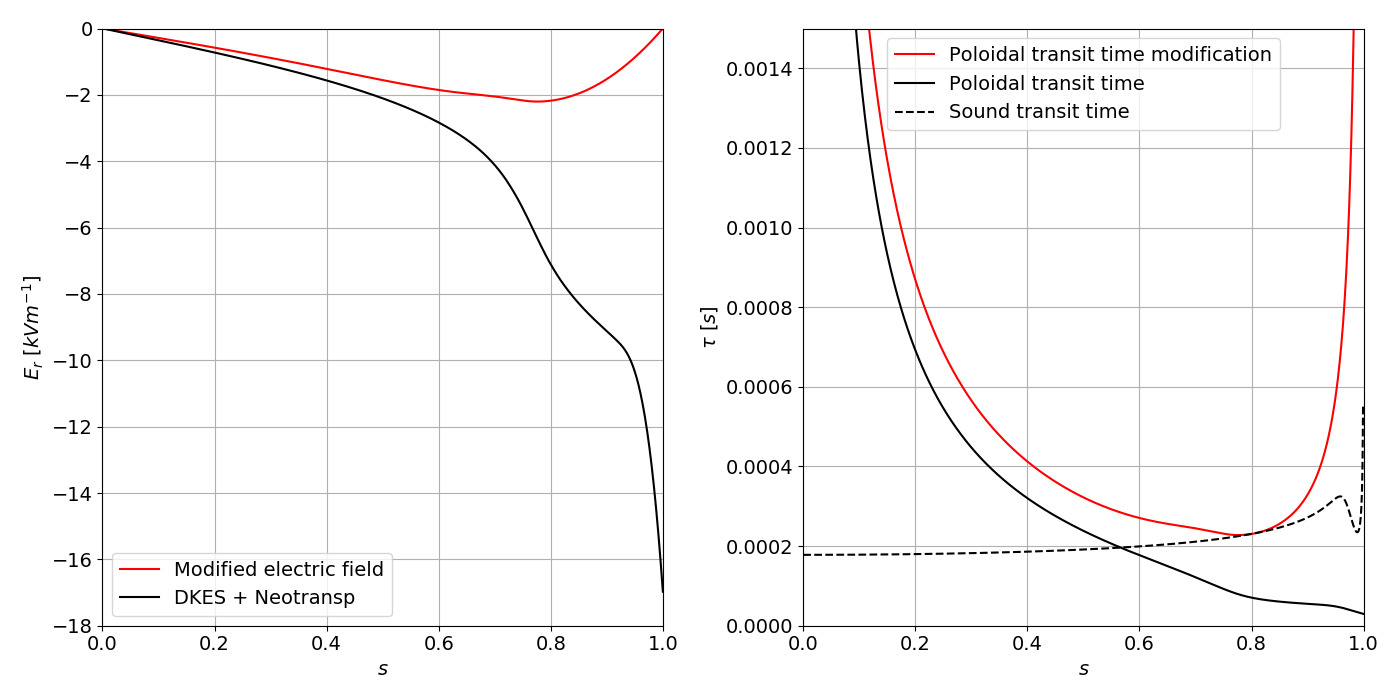}
    \caption{Computed and modified radial electric field (a) from neoclassical transport calculations. Comparing the approximate timescales of poloidal and parallel flow dynamics (b), the computed radial electric field implies the formation of shocks in the plasma periphery. The modified profile remains subsonic.}
    \label{fig:w7as_e_field}
\end{figure}

The calculated electric field is shown in Figure \ref{fig:w7as_e_field}. It can be seen that the approximate sound transit time is exceeded by the poloidal transit time of the background $\mathbf{E} \times \mathbf{B}$ flows, where these two timescales are given by

\begin{equation}
    \tau_{s} = \frac{2 \pi R_0 q}{c_s},
\end{equation}

and

\begin{equation}
    \tau_{\mathbf{E}\times\mathbf{B}} = \frac{\textcolor{black}{B_0^2} 2 \pi <r_{minor}>(s)}{<E_r>(s) \times B_0},
\end{equation}

where $c_s$ is the sound speed and $<a>$ refers to the flux surface average of $a$. The condition $\tau_s > \tau_{\mathbf{E}\times\mathbf{B}}$ implies \textcolor{black}{that the equilibrium $\mathbf{E}\times\mathbf{B}$ flow can lead to shocks, as observed in the literature for ideal MHD equilibria with flows in tokamaks \citep{betti2000radial}}. For a valid equilibrium condition, the equilibrium flows must remain subsonic. In addition, it is expected that the electric field should become more radially outward as the separatrix is approached. This is because in the open field line region outside the plasma, the electrons are expected to free stream faster towards the divertor targets, leading to an outward electric field. It should be noted that the field outside the confined region is only approximately radial as it is induced by open field line transport along fieldlines, but the transition towards this condition should be reflected in the $E_r$ profile close to $s=1$. The profile is therefore modified in order to ensure that the poloidal flow is subsonic over the domain, and the electric field transitions to zero at the plasma boundary. It should be acknowledged that this introduces an inconsistency between the neoclassical prediction of the radial electric field, and that enforced in simulations. This inconsistency is considered to be justified given the relatively high collisionality of the plasma, and uncertainties in the true experimental profiles for density and temperature. If more accurate predictions of the experimental profiles were available, it may be that the corresponding $E_r$ would be in better agreement with the red line in Figure \ref{fig:w7as_e_field} (a).

The main simulation parameters are listed in Appendix \ref{app:params}. For this case, these parameters are intended to approach those of the experiment for comparison. The plasma resistivity and perpendicular viscosity have a Spitzer-like $T^{-3/2}$ dependence. Compared to the simulations in \citep{ramasamy2024nonlinear}, the perpendicular particle and heat diffusivities have been increased to better approximate the value of the experiment. The parallel heat conductivity is kept constant, as this is found to improve numerical stability in the nonlinear phase. Similar to the W7-X-like case, $n_{tor}$ has been kept below 30 to reduce the computational expense of simulations. Past results in \citep{ramasamy2024nonlinear} have shown that  higher toroidal harmonics can have an influence on the detailed nonlinear dynamics, at least in the absence of flows. A simulation has been carried out in the absence of background flows with $n_{tor} \le 50$ which also shows the presence of such low n modes (not shown). As this study pursued the low n modes observed in the experiment, the choice of resolution is not expected to have an impact on the results.

It is important to note the set up of the heat source, $S_E$, in equation \ref{eq:temp_equation}. The heat source is localised in the plasma core using the summation of two gaussian distributions with variance $s=0.125$ and $0.33$, scaling the profile to the specified heating power. The concentration of the heat source in the plasma core is intended to be comparable to the heat deposition of the co-directional NBI current drive, which induces more significant currents on axis \citep{geiger2003analysis}. 

At the same time, this heat source does not maintain the initial profiles against the diffusive transport, causing the plasma pressure to redistribute with a higher peaking factor in the centre of the device. Large deviations from the initial profiles can lead to lower fidelity dynamics, as a strong shift of the flux surfaces can mean that the driven poloidal flows are no longer perfectly aligned along flux surfaces. The advantage of this approach however, is that it allows one to see if the plasma column can be maintained at the prescribed experimental heating power, and if the modification of the background plasma profile, which is artificially prescribed as linear, has an influence on the dynamics. As shown in Appendix \ref{app:reduced_mhd}, the background profiles can also be maintained in time against the diffusive operators. A simulation using this approach is shown in Section \ref{sec:w7as_nonlin}, for comparison.

\subsection{Nonlinear dynamics} \label{sec:w7as_nonlin}
\begin{figure}
    \centering
    \includegraphics[width=0.55\linewidth]{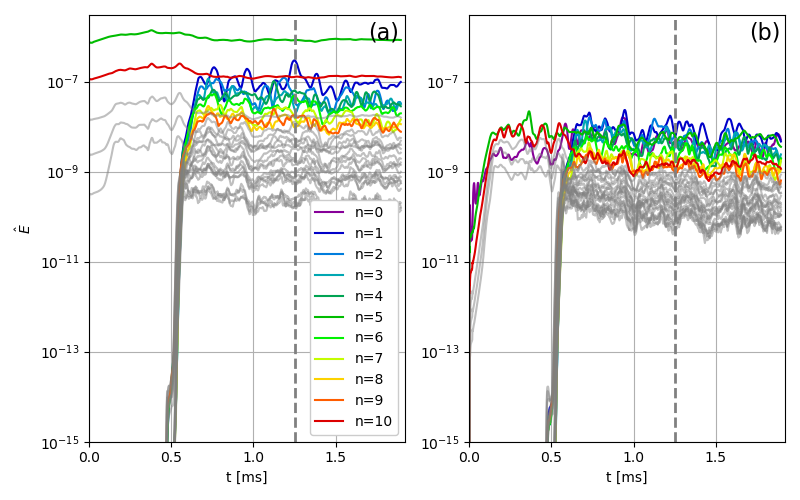}
    \includegraphics[width=0.42\linewidth]{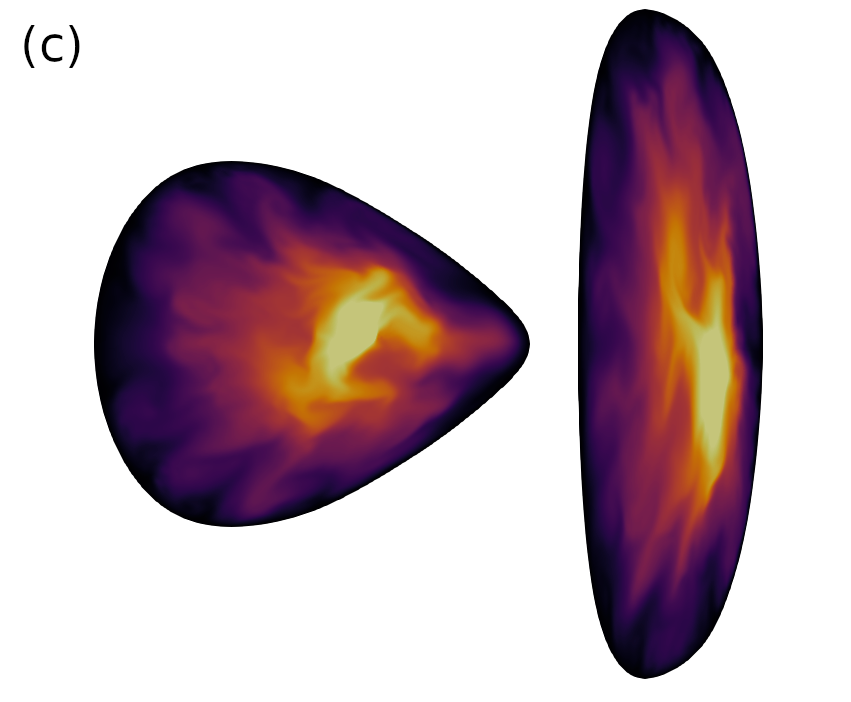}
    \includegraphics[width=0.55\linewidth]{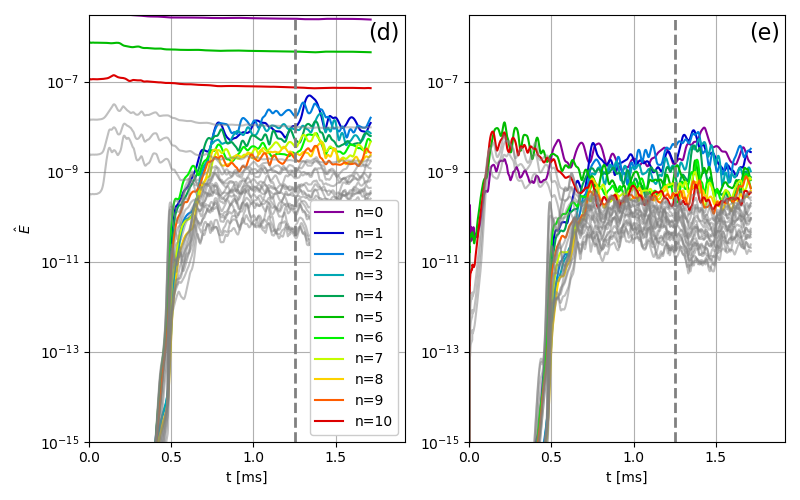}
    \includegraphics[width=0.42\linewidth]{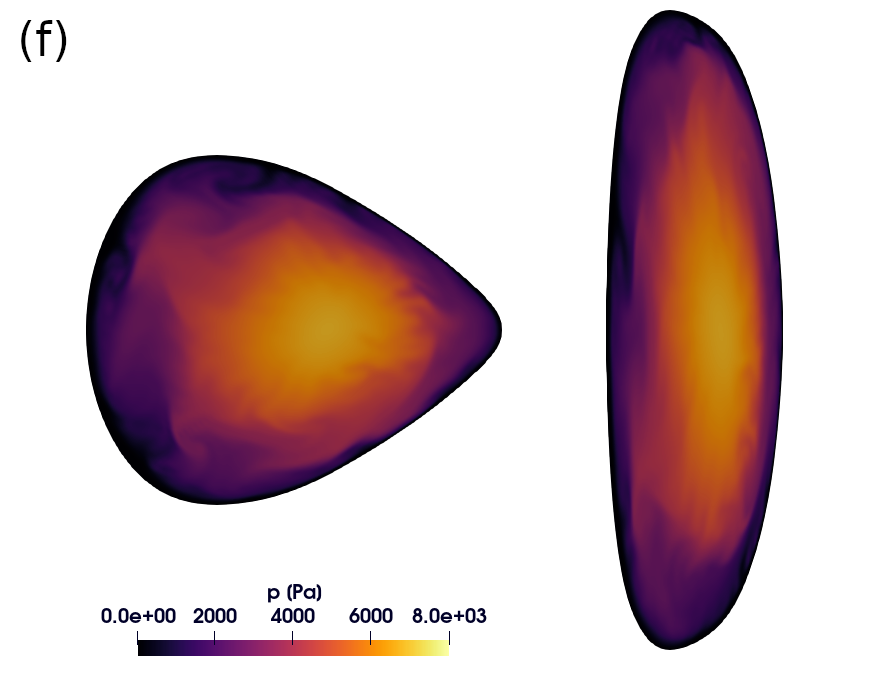}
    \caption{Magnetic (a, d) and kinetic (b, e) energies of W7-AS cases without flows using a gaussian heating profile (top row) and artificially maintaining the initial equilibrium profiles (bottom row). \textcolor{black}{The pressure is shown in the $\phi=0$ and $\phi=\pi$ poloidal planes (c) and (f) at the time point corresponding to the grey dashed line.} In the case with a gaussian heat source (top row), the $n=1$ magnetic energy is larger than other mode numbers belonging to the $N_f > 0$ mode families, the energy signature is not sustained, and the case is unstable to multiple, overlapping modes. At $t=1.25\ ms$, a clear (1, 2) mode structure is present in the pressure (top, right). In the case with a maintained equilibrium profile (bottom row), a weaker low $n$ magnetic energy signature is observed, which corresponds to the low $n$ perturbation being restricted to the plasma outer mid-radius.}
    \label{fig:w7as_low_n_modes}
\end{figure}

The dynamics of the outlined W7-AS equilibrium in the absence of background equilibrium flows is first considered in Figure \ref{fig:w7as_low_n_modes}. Two simulations are carried out using the different heating methods discussed in Section \ref{sec:w7as_setup}. The simulations are initialised first with only the $N_f=0$ mode family. Once these modes are saturated, the $N_f>0$ modes are then introduced and rapidly grow from noise level to compete with the equilibrium modes. 

Consider first the case using a gaussian heat source in the top row of Figure \ref{fig:w7as_low_n_modes}. Once initialised, the saturation of the $N_f>0$ mode families is led by high n modes which saturate at modest values around $t=0.6\ ms$. This is partly due to the competition with the already nonlinearly saturated $N_f=0$ modes. It is only later in time that the low n modes grow to comparable kinetic energy values to that of the equilibrium modes. In the magnetic energies, a $n=1$ signature is observed that remains dominant over the other $N_f>0$ modes. Plotting the pressure in the $\phi=0.0^\circ$ and $180.0^\circ$ planes at $t=1.25\ ms$, a clear (1, 2) mode structure can be seen. This structure is not consistently maintained over the nonlinear phase, however, as it competes with a mix of other instabilities. 

In the bottom row of Figure \ref{fig:w7as_low_n_modes}, a simulation is shown, where the initial equilibrium profile is maintained by subtracting the equilibrium profiles from the diffusive operators in equation \ref{eq:density_equation} and \ref{eq:temp_equation}. In this simulation, weaker magnetic energy signatures corresponding to the $n=1$ and $n=2$ modes are observed. The plots of the pressure distribution of this case indicate that low mode number poloidal structures are relatively faint and restricted more towards the outer mid-radius of the plasma. As discussed in Section \ref{sec:w7as_setup}, the suppression of low mode number instabilities could indicate that the assumed linear profile is not representative of the true experimental conditions, which could have a more peaked profile.

\begin{figure}
    \centering
    \includegraphics[width=0.55\linewidth]{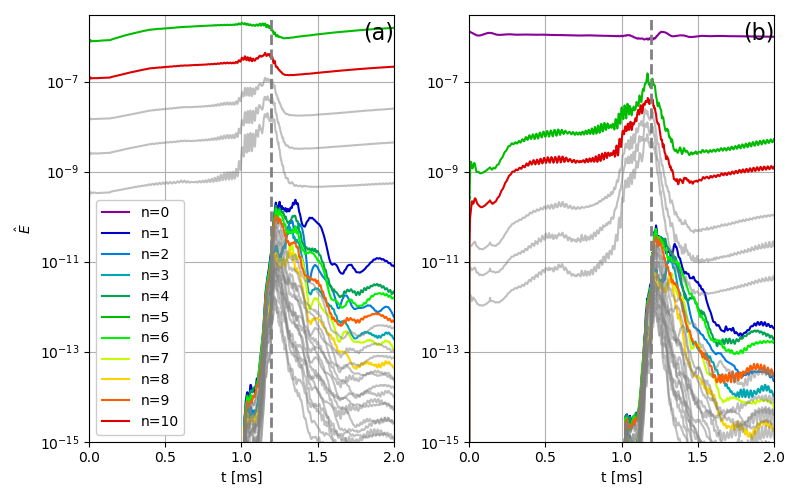}
    \includegraphics[width=0.44\linewidth]{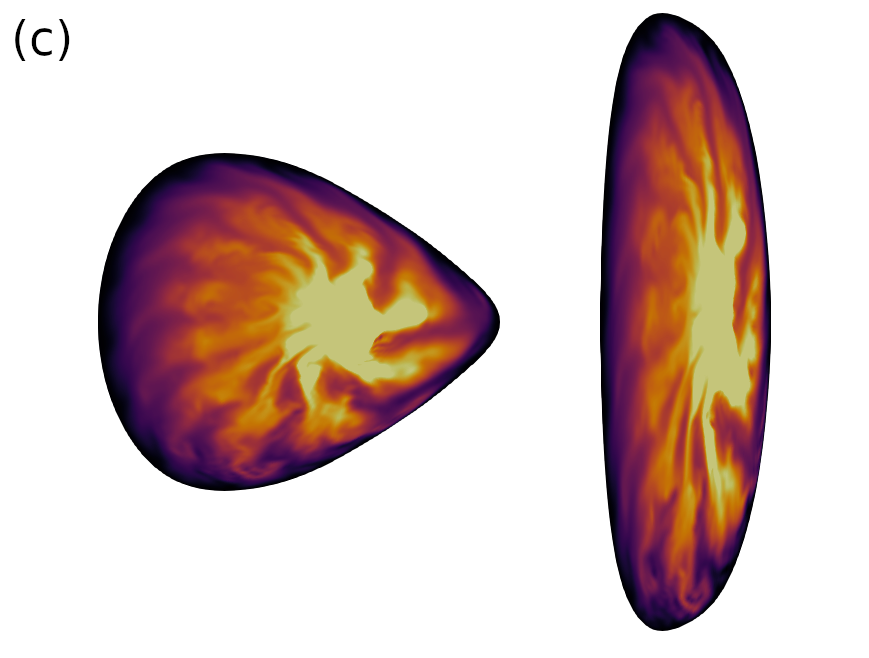}
    \includegraphics[width=0.55\linewidth]{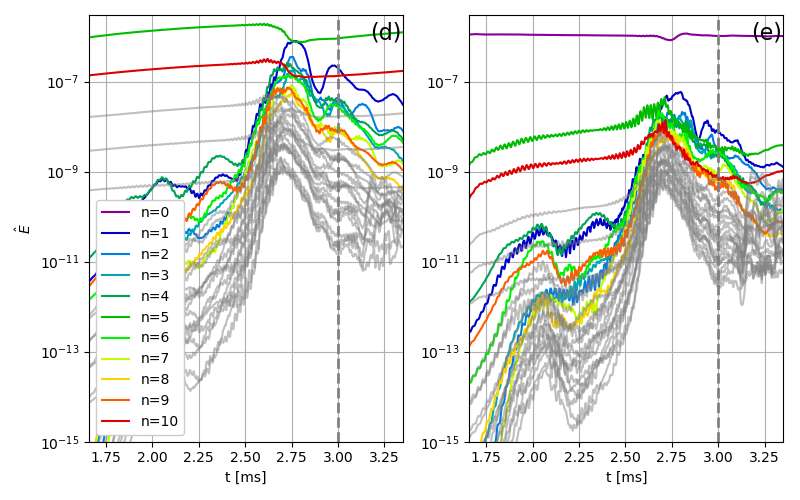}
    \includegraphics[width=0.44\linewidth]{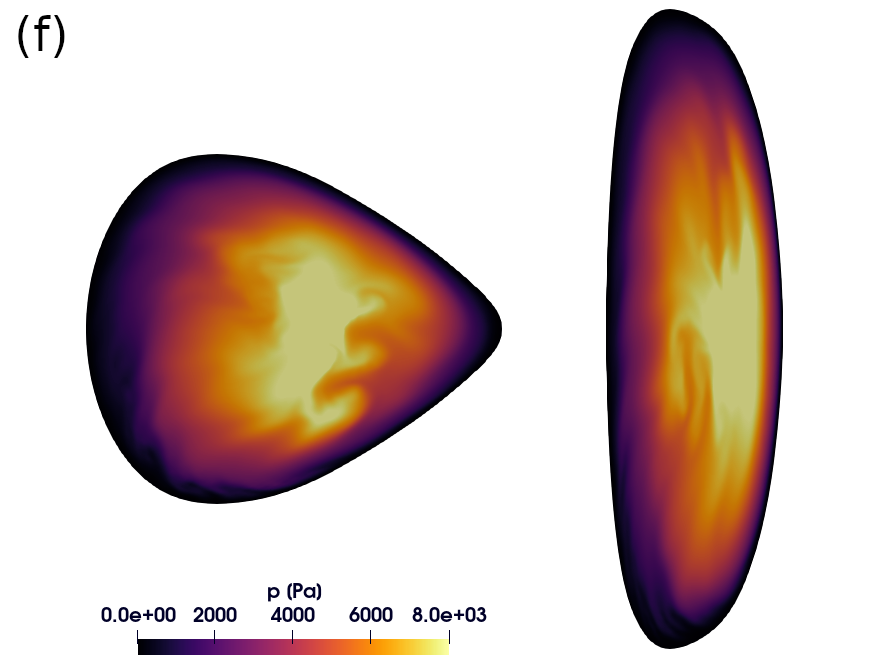}
    \caption{Magnetic (a, d) and kinetic (b, e) energies of W7-AS case with the background flow profile shown in Figure \ref{fig:w7as_e_field}. \textcolor{black}{The pressure is shown in the $\phi=0$ and $\phi=\pi$ poloidal planes (c) and (f) at the time point corresponding to the grey dashed line.} In the case where the $N_f=0$ modes evolved for longer, prior to initialising the $N_f > 0$ mode families (top row), it can be seen $t<0.8 ms$ that the high toroidal modes have significantly less energy than the equivalent case without flows in the top row of Figure \ref{fig:w7as_low_n_modes}. This allows the $n=5$ mode to dominate over the whole plasma column leading to a transient crash at $t \approx 1.15\ ms$. If the $N_f=1$ mode family is allowed to grow in the early linear phase (bottom row), such that it can compete with the $N_f=0$ mode family, a dominant (1, 2) mode is observed as in the experiment.}
    \label{fig:w7as_low_n_modes_with_flows}
\end{figure}

Simulations, which include the presence of a background equilibrium flow are shown in Figure \ref{fig:w7as_low_n_modes_with_flows}. The initial dynamics are quite different from the stationary equilibria in Figure \ref{fig:w7as_low_n_modes}. Early in the simulation, before the $N_f>0$ modes are initialised, the high n modes belonging to the $N_f=0$ mode family grow more slowly than in the case without flows, where the nonlinear saturation and stochastisation of the outer plasma is rapid, occurring within the first $t<0.15\ ms$ of the simulation. This behaviour is expected given the equilibrium flow shear will preferentially stabilise higher mode numbers.

Later in the nonlinear phase, the absence of higher n modes that compete with the $n=5$ mode allows a stronger crash to occur. This is because the higher n modes contribute to enhanced transport, preventing the pressure from building up in the core. The $n=5$ mode responds to this build up of pressure from the applied heat source, crashing with a higher kinetic energy at $t \approx 1.2\ ms$ than in the absence of flows. After the crash, the plasma recovers, and pressure starts to build up again.

The observed $n=5$ mode is not the expected dominant MHD signature from the experiment. To test the sensitivity of the result on the initial conditions, the $N_f=1$ mode family is allowed to grow in the initial phase of the simulation, in a similar way to the analysis carried out in Section \ref{sec:w7x}. This simulation is shown in the bottom row of Figure \ref{fig:w7as_low_n_modes_with_flows}. It can be seen that the $N_f=0$ and $2$ mode families are allowed to evolve while the $N_f=1$ mode family is still subdominant. The subsequent evolution of the $N_f=1$ mode family shows that the $n=4$ mode saturates first at a subdominant value, before the $n=1$ mode overtakes it and takes part in a crash in combination with the $n=5$, $n=2$ modes. This initial crash occurs at a higher energy than the equivalent simulation without flows in Figure \ref{fig:w7as_low_n_modes}, again because of the build up in the core pressure due to the heat source. Later in time at $t=3\ ms$, a dominant (1, 2) mode structure is again observed.   

\begin{figure}
    \centering
    \includegraphics[width=\linewidth]{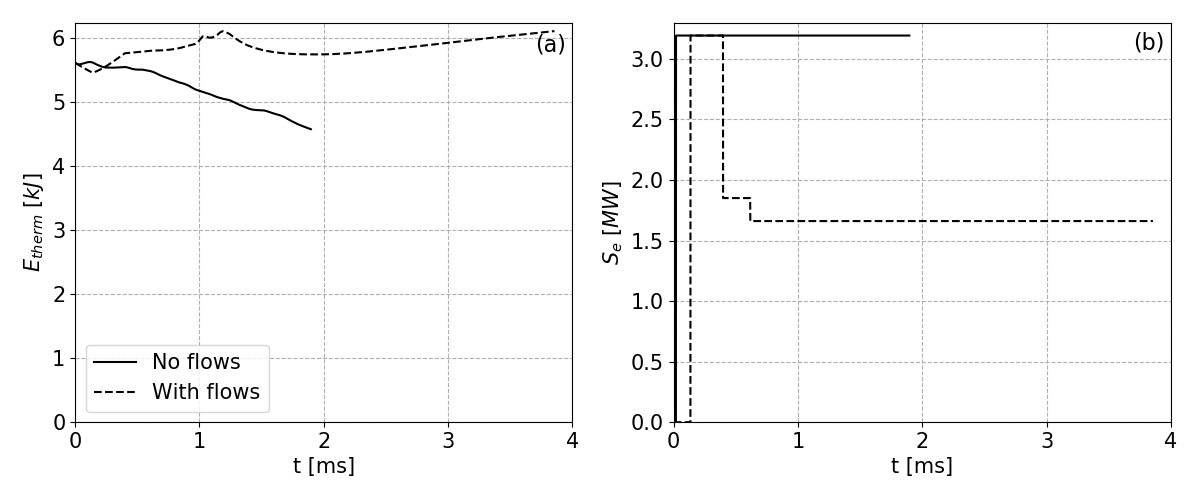}
    \caption{Integrated thermal energy over the plasma volume (a) and applied heating power (b) in cases with (dashed) and without (solid) flows. Without flows, the degradation in confinement is too significant to sustain the plasma at the experimental heat source. With flows, the plasma can be sustained at $1.6\ MW$ of heating power, within the range of the experiment.}
    \label{fig:w7as_therm}
\end{figure}

Beyond the observation of the (1, 2) mode, it is also important to determine whether the transport properties are consistent with the experiment. Figure \ref{fig:w7as_therm} shows that the evolution of the total thermal energy in simulations with and without flows in the top rows of Figure \ref{fig:w7as_low_n_modes} and \ref{fig:w7as_low_n_modes_with_flows}, respectively. The total applied heating power is also shown, which is varied in time, within the limit of the maximum experimental heat source, $3.2\ MW$. When interpreting this result, it is important to note that the applied core localised heating source is a best case scenario for confining the applied thermal energy, because the heat must cross the whole plasma volume to escape the domain. In the absence of flows, the initial thermal energy cannot be maintained\textcolor{black}{, due to enhanced convective transport}. The introduction of equilibrium flows sufficiently stabilises the MHD activity, such that the thermal energy can be sustained with reasonable thermal diffusivities, and heating below $3.2\ MW$. In such a way, it could be argued that the equilibrium flows were a necessary ingredient in maintaining the plasma confinement in W7-AS.

\section{Discussion on benign nonlinear saturation of Mercier unstable modes} \label{sec:discuss}
The analysis in Section \ref{sec:w7x} and \ref{sec:w7as} has been carried out with the primary goal of understanding nonlinear saturation of low n pressure driven modes in Wendelstein stellarators. In particular, how have low n pressure driven modes saturated benignly in W7-AS, and could similar behaviour be observed in W7-X? Evidence that flux pumping scenarios in tokamaks correspond to the nonlinear saturation of a Mercier unstable initial equilibrium state has been provided in Section \ref{sec:intro}. For this reason, this study also aims to interpret the nonlinear stability of Mercier unstable configurations through the lens of the flux pumping tokamak literature. Can a similar flux pumping scenario be found in stellarators? The above questions could only be partially answered within the current simulation capabilities, time frame and computational resource available for this study. Below an attempt at navigating the current progress is made, in order to inform the next steps.

In the simulated W7-X-like case, a clear MHD dynamo term is observed in the initial saturation of a (2, 3) interchange mode. This observation provides a strong indication that there must be a link to the nonlinear stability of such modes and flux pumping theory. The immediate problem with applying the standard flux pumping theory outlined in Appendix \ref{app:flux_pumping} to Wendelstein 7-X, and QI stellarators more generally, is that the net toroidal current is negligibly small. Flux pumping typically requires a redistribution of the plasma current density to counterbalance the MHD dynamo on a resistive time scale. Without a plasma current source, it is unclear how the dynamo induced by the MHD activity can be sustained long term. In the simulations presented in Section \ref{sec:w7x}, the MHD dynamo cannot be followed over a resistive timescale to interrogate this question, because a crash-like behaviour of the core temperature and partial reconnection is observed after the initial nonlinear saturation. This behaviour is not surprising given the simulated case is far from marginality. The authors would argue that the partial reconnection observed in the simulated case has similarities to the crash dynamics proposed in \citep{itoh1995sawtooth, jardin2020new} for sawtooth crashes in tokamaks. Such crash dynamics could also apply to stellarators.

A natural next step for this case would be to simulate similar configurations closer to marginality, in an attempt to observe a flux pumping scenario. Such simulations would be interesting to carry out in order to understand if the plasma can self-organise to a consistent MHD dynamo in the absence of \textcolor{black}{an externally driven current source, where only Pfirsch-Schl\"uter currents are significant. If Pfirsch-Schl\"uter currents cannot compete against the MHD dynamo, a modest amount of current drive can be applied to maintain the mode close to marginality, in order to maintain good confinement, and hopefully reach higher performance, similar to \citep{burckhart2023experimental}.} Given the uncertainties raised regarding the exact conditions for observing such a nonlinear saturation, that no experimental observations of such scenarios have been observed in W7-X yet to inform simulations, and that simulations close to marginality are very expensive with current model performance, it is currently not clear how to proceed efficiently towards such a goal.

At this stage, the authors chose to focus current efforts on reproducing past experimental observations of low n saturated modes in W7-AS. In Section \ref{sec:w7as}, low n MHD activity has been reproduced at the approximate experimental plasma resistivity and higher more realistic thermal diffusivity, going beyond previous attempts in \citep{ramasamy2024nonlinear}. At the same time, the (1, 2) mode is transient in simulations. The $\mathbf{E}\times\mathbf{B}$ flow has been included in simulations in order to stabilise the high n mode branch which should suppress the crash dynamics due to the overlap of multiple MHD modes. While the introduced flow shear is shown to have this effect, multiple MHD modes are still unstable, and it is found that the (5, 10) mode dominates nonlinearly in simulations. If the $N_f=1$ mode family is allowed to grow prior to initialising the $N_f=0$ mode family, the (1, 2) mode is observed again, indicating the dynamics are sensitive to the initial conditions.

A coherent sustained (1, 2) mode like the one observed in the experiment has not been found in any of the simulations carried out herein. It should be acknowledged that the validation against such a past decomissioned experiment is challenging. The experimental equilibrium reconstructions contain many assumptions - most prominently the assumption of an initially linear pressure profile $p(s)=p_0 (1 - s)$. This assumption is likely incorrect given observations of peaked profiles in experimental diagnostics of similar discharges \citep{zarnstorff2005equilibrium}. Results in Section \ref{sec:w7as_nonlin} show that when the pressure profile becomes more peaked, the MHD signature of the low n modes is more prominent, indicating that perhaps different assumptions on the equilibrium profiles would make the (1, 2) mode easier to observe. Exploring the impact of the pressure profile would be more appropriate with a linear viscoresistive MHD code like CASTOR3D. Initial attempts at carrying out such analysis proved numerically challenging and are reserved for future work.  

In addition to these equilibrium reconstruction uncertainties, it is not clear to the authors whether the strong $n=2$ and $3$ signatures in the Mirnov coil diagnostic at approximately 20 and 30 kHz in Figure 6 of \citep{weller2003investigation} are sidebands of the dominant $n=1$ harmonic, or separate competing modes. These uncertainties make it difficult to be sure that a coherent (1, 2) mode should be observed with the current simulation set up.

Lastly, it could be argued that the discrepancy between simulation and experiment may have a link to flux pumping. In particular, the equilibrium reconstructions used herein have assumed that there are no internal current sources in the simulation. Current control is required in these high $\beta$ discharges, in order to suppress NBI driven currents \citep{geiger2003analysis}. Internal currents can significantly change the internal $\iota$ profile, modifying the MHD stability. They would also provide a counterbalance to the MHD dynamo produced by MHD instabilities, such that \textcolor{black}{the (1, 2) mode observed in simulations can be sustained, as the mode remains close to marginality, avoiding transient dynamics. This explanation would be consistent with a typical flux pumping scenario, however it should only be considered a hypothesis at this stage, which could warrant further investigation to validate. Such an effort would be better informed first by linear MHD analysis.}

\section{Conclusion} \label{sec:conclusion}

The above analysis has focused on low n interchange instabilities, intending to shed light on the dynamics of Mercier unstable Wendelstein stellarators by comparing the observed dynamics to those of tokamak flux pumping scenarios. The overarching goal of these nonlinear simulations is to provide insights into how one might operate stellarators in regimes where Mercier stability is violated. While the MHD dynamo produced in the early nonlinear phase of the dynamics has the same characteristic features of marginal quasi-interchange modes in tokamaks, implying flux pumping may be possible, partial reconnection and crash dynamics are observed in simulations later in time. \textcolor{black}{This could be because the simulations lack typical ingredients of flux pumping scenarios, namely that they are (1) initialised too far from marginality to observe nonlinear saturation, and (2) without a current source or control mechanism, which is a necessary ingredient for the observation of flux pumping in tokamaks.}

The results herein are presented as a first step, and there are many directions for future studies to attack. Regarding the study of low n interchange instabilities in W7-X, \textcolor{black}{the focus of the present study has been to provide an initial theoretical understanding of how low n modes may saturate benignly.} It would be interesting to consider an experimentally relevant configuration such that nonlinear results can be compared and grounded by real world measurements. In particular, a root to realising the envisaged flux pumping scenario in a stellarator could be to exploit a NBI or ECCD current source with an appropriate deposition profile, similar to the strategy achieved on ASDEX Upgrade \citep{burckhart2023experimental}. \textcolor{black}{As part of this work, it will be interesting to understand the importance of the high $n$ mode branch, and its potential to modify the detailed nonlinear dynamics. In the absence of finite larmor radius effects and other stabilisation mechanisms, there is likely to be competition and interplay between these modes, similar to past observations in LHD discharges \citep{ohdachi2010density, varela2011ballooning}, which can change the observed experimental dynamics.} Such experimental discharges are not available at the time of writing, but should hopefully be carried out in some form as part of the W7-X experimental program.

In W7-AS, \textcolor{black}{while present results have improved the qualitative agreement with the experiment by removing the high n mode branch, further work is necessary to robustly reproduce the experimental low n mode activity and understand whether there is a link between these real world observations and flux pumping theory}. Several strategies could improve the match between experiment and simulation. Among the limitations in the modeling, it should be noted that the simulations have been conducted from an initial condition which is already MHD unstable. It would be interesting to follow the approach used in other studies \citep{wright2024mhd, cathey2022mhd}, where simulations transition across the stability boundary. For the Shafranov shift and change in the equilibrium with increasing $\beta$ to be appropriately accounted for in such simulations, the divertor region must be appropriately modeled at finite $\beta$, perhaps also removing the constraint of a fixed boundary condition. Such a scrape off layer extension of JOREK is in development.

Independent of device specific analysis, it would be useful to interrogate the flux pumping mechanism further in stellarators. In particular, tokamak simulations seem to observe flux pumping when a single low n mode is unstable \citep{zhang2024full, jardin2020new}. It would therefore be interesting to search the linear parameter space of such devices to find such a regime, and test if the crash behaviour observed in the present study can be removed when the threat from overlapping modes is eliminated. 

Lastly, \textcolor{black}{it would be interesting to consider whether the discussed mechanisms play a role in the nonlinear saturation of low n interchange modes observed experimentally in inward shifted LHD configurations \citep{sakakibara2002effect}}, or Mercier unstable quasi-axisymmetric stellarators. In particular, quasi-axisymmetric stellarators are a more natural analog to tokamaks and are known to have a significant bootstrap current. As shown in \citep{yu2024numerical}, the bootstrap current can play an important role in sustaining an MHD dynamo, such that the path towards a flux pumping regime is clearer. 

\section*{Acknowledgements}
The authors would like to thank Matthias Hoelzl, Ksenia Aleynikova, Nikita Nikulsin, Alessandro Zocco, and Sibylle G\"unter for helpful discussions. The simulations presented in this work were carried out on the Gauss Centre for Supercomputing cluster, HAWK, in Germany, and the EUROfusion High Performance Computer (Marconi-Fusion).

\section*{Funding}
This work has been carried out as part of a EUROfusion Bernard Bigot Researcher Grant within the framework of the EUROfusion Consortium, funded by the European Union via the Euratom Research and Training Programme (Grant Agreement No 101052200 — EUROfusion). Views and opinions expressed are however those of the author(s) only and do not necessarily reflect those of the European Union or the European Commission. Neither the European Union nor the European Commission can be held responsible for them.

\section*{Declaration of interests}
The authors report not conflict of interests.

\appendix

\section{Review of Flux pumping theory applied to a no net toroidal current carrying stellarator} \label{app:flux_pumping}

This section reviews past flux pumping theory, applying it to the stellarator context. The analysis here follows that found in \citep{krebs2017magnetic}. Nonlinear saturation requires a steady state solution to the MHD equations. This implies that finite flows induced by MHD modes are constantly redistributing the injected thermal energy and helicity from external heating and current sources. In the case of stellarators, there is normally no loop voltage applied to the plasma, but external and internal current sources can still play a role. To understand flux pumping, consider the induction equation

\begin{equation}
    \frac{\partial \mathbf{A}}{\partial t} = - \eta_{neo} (\mathbf{j} - \mathbf{j}_{CD} - \mathbf{j}_{BS}) + \left(\mathbf{v} \times \mathbf{B}\right) - \nabla \Phi,
    \label{eq:induction}
\end{equation}

where $\mathbf{j}$ is the total current, $\mathbf{j}_{CD}$ is the externally driven current sources from neutral beam injection (NBI) and electron cyclotron current drive (ECCD), and $\mathbf{j}_{BS}$ is the internal current source from bootstrap current. Note that these sources are not included in the model outlined in \citep{krebs2017magnetic}, because the changes in the current profile are induced by heating, which modifies the peaking of the temperature profile. 

In the case of a tokamak, equation \ref{eq:induction} is normally split into axisymmetric and $n=1$ components at this point. In the case of a stellarator, the induction equation must instead be split between the equilibrium, $N_f=0$, and periodicity breaking, $N_f>0$, mode families. The $N_f=0$ modes are assumed to approximately correspond to the time evolving equilibrium quantities. As any background equilibrium flow along magnetic surfaces will not contribute a change to the magnetic field, this component of the $\mathbf{v} \times \mathbf{B}$ component is neglected. The $N_f=0$ component of the induction equation then becomes

\begin{equation}
    \frac{\partial \mathbf{A}_{N_f=0}}{\partial t} = -[\eta_{neo} (\mathbf{j} - \mathbf{j}_{CD} - \mathbf{j}_{BS}) ]_{N_f=0} - [\eta_{neo} (\mathbf{j} - \mathbf{j}_{CD} - \mathbf{j}_{BS})]_{N_f>0} +  [\mathbf{v} \times \mathbf{B}]_{N_f>0}.
\end{equation}

Finally, assuming a steady state solution, separating the $N_f=0$ terms into equilibrium, $A_{equil}$ and perturbed quantities, $\Delta A$, and neglecting second order perturbed quantities, we arrive at

\begin{eqnarray}
    0 & = & -\eta_{equil} (\Delta \mathbf{j} - \Delta \mathbf{j}_{CD} - \Delta \mathbf{j}_{BS}) - \Delta \eta (\mathbf{j}_{equil} - \mathbf{j}_{CD, equil} - \mathbf{j}_{BS, equil}) \nonumber \\
    && - [\eta_{neo} (\mathbf{j} - \mathbf{j}_{CD} - \mathbf{j}_{BS})]_{N_f>0} + [\mathbf{v} \times \mathbf{B}]_{N_f>0},
\end{eqnarray}

where the symmetry breaking $N_f>0$ resistive term has been neglected because it is found to be negligible in past simulation studies. The structure of this equation is very similar to that in \citep{krebs2017magnetic}. The crucial difference for flux pumping in the context of stellarators is not so much these equations, as the nature of the currents in such devices. The induced current in a QI stellarator is typically optimised to be zero. This implies that there is no term in the standard stellarator picture that can counterbalance the perturbed $\mathbf{v}\times\mathbf{B}$ term. One possibility to achieve flux pumping would be through external current drive. If ECCD or NBI currents are driven to counteract the dynamo voltage induced by a perturbation, it is possible that the MHD activity would saturate. For QS stellarators, the bootstrap current is a more promising alternative. A recent study has shown that the bootstrap current can play an important role in the formation of flux pumping scenarios in tokamaks \citep{yu2024numerical}.

\section{Nonlinear MHD model} \label{app:reduced_mhd}
The main simulations carried out in this study make use of the stellarator capable reduced MHD model originally derived in \citep{nikulsin2021models}. In this model, the single fluid viscoresistive MHD equations are solved for using a reduced MHD ansatz, such that $\mathbf{B} = \nabla \Psi \times \nabla \chi + \nabla \chi$ and $\mathbf{v}=\frac{\nabla \Phi \times \nabla \chi}{B_v^2} + v_\parallel \mathbf{B}$. This yields the following equations, which are evolved in time

\begin{eqnarray} \label{eq:induction_equation}
        \frac{\partial \Psi}{\partial t} &= \frac{\partial^\parallel \Phi - \left[\Psi, \Phi\right]}{B_\mathrm{v}}  + \eta \left(j - j_\mathrm{source}\right) - \nabla \cdot \left(\eta_\mathrm{num} \nabla^\bot j \right)
\end{eqnarray}

\begin{eqnarray} \label{eq:vorticity_equation}
    \nabla \cdot \left( \frac{\partial}{\partial t}\left[\frac{\rho}{B_\mathrm{v}^2}\nabla^\bot \Phi \right] \right) & =  &\frac{B_\mathrm{v}}{2} \left[ \frac{\rho}{B_\mathrm{v}^2}, \frac{\left(\Phi, \Phi\right)}{B_\mathrm{v}^2} \right] - B_\mathrm{v} \left[ \frac{\rho \omega}{B_\mathrm{v}^4}, \Phi \right]  - \nabla \cdot \left[ \frac{\nabla^\bot \Phi}{B_\mathrm{v}^2}  \nabla \cdot \left(\rho \mathbf{v}\right) \right] \nonumber\\
                                && + \nabla \cdot \left(\left[\frac{\nu \rho}{B_\mathrm{v}^2}\nabla^\bot (\Phi - \Phi_0) \right] \right)\nonumber\\
                                && + \nabla \cdot \left( j \mathbf{B} \right) + B_\mathrm{v} \left[ \frac{1}{B_\mathrm{v}^2}, p \right]+ \nabla \cdot \left( \mu_\bot \nabla^\bot \omega \right) \nonumber\\
                                &&- \Delta ^\bot \left( \mu_\mathrm{num} \Delta^\bot \omega \right)
\end{eqnarray}

\begin{eqnarray} \label{eq:density_equation}
    \frac{\partial \rho}{\partial t} = &-B_\mathrm{v} \left[ \frac{\rho}{B_\mathrm{v}^2}, \Phi \right] - B_\mathrm{v} \partial^\parallel (\rho \mathrm{v}_\parallel) - B_\mathrm{v} \left[\rho \mathrm{v}_\parallel, \Psi \right] + P
\end{eqnarray}

\begin{eqnarray} \label{eq:temp_equation}
        \frac{\partial \left(\rho T\right)}{\partial t} & = &	-\frac{1}{B_\mathrm{v}}\left[ \rho T, \Phi \right] - \mathrm{v}_\parallel B_\mathrm{v} \partial^\parallel p - \mathrm{v}_\parallel B_\mathrm{v} \left[p, \Psi \right] - \Gamma \rho T B_\mathrm{v} \left[ \frac{1}{B_\mathrm{v}^2}, \Phi\right] \nonumber\\
                                             &&  - \Gamma p B_\mathrm{v} \partial^\parallel \mathrm{v}_\parallel - \Gamma p B_\mathrm{v} \left[ \mathrm{v}_\parallel, \Psi \right] \nonumber\\                                   
                                             &&  + \nabla \cdot \Bigg[ \Bigg. \kappa_\bot \nabla_\bot (T - T_0) + \kappa_\parallel  \nabla_\parallel (T - T_0) + T D_\perp \nabla_\bot (\rho - \rho_0) +T D_\parallel \nabla_\parallel (\rho - \rho_0)  \Bigg.\Bigg] \nonumber\\
                                             && + \left( S_e + \eta_\mathrm{Ohm} (\Gamma - 1) B_\mathrm{v}^2 j^2\right)
\end{eqnarray}

\begin{eqnarray} \label{eq:vpar_equation}
    B^2 \frac{\partial \left(\rho \mathrm{v}_\parallel \right)}{\partial t} + \frac{\rho \mathrm{v}_\parallel}{2}\frac{\partial B^2}{\partial t} & = & -\frac{\rho B_\mathrm{v}}{2}\partial^\parallel \mathrm{v}^2 - \frac{\rho B_\mathrm{v}}{2} \left[\mathrm{v}^2, \Psi \right]  - \mathrm{v}_\parallel B^2 \nabla \cdot \left( \rho \mathbf{v} \right) \nonumber\\
                                && - B_\mathrm{v} \partial^\parallel p - B_\mathrm{v} \left[p, \Psi\right] \nonumber\\
                                && + B^2 \nabla \cdot \left[\mu_{\parallel, \bot} \nabla_\bot \mathrm{v}_\parallel +\mu_{\parallel, \parallel} \nabla_\parallel \mathrm{v}_\parallel\right]
\end{eqnarray}

where the auxiliary variables for the plasma current, $j$, and vorticity, $\omega$ are defined as

\begin{equation}
    j = \Delta^* \Psi
\end{equation}

\begin{equation}
    \omega = \Delta^\bot \Phi
\end{equation}

and $P = \nabla \cdot \left(D_\bot \nabla_\bot (\rho - \rho_0) + D_\parallel \nabla_\parallel (\rho - \rho_0) \right)+ S_\rho$. In the above equations, the operators are defined as follows

\begin{eqnarray*}
\begin{array}{cc}
\nabla_\parallel = B^{-2} \mathbf{B}(\mathbf{B}\cdot \nabla, & \nabla_\bot = \nabla - \nabla_\parallel \\
\partial^\parallel = B_\mathrm{v}^{-1} \nabla \chi \cdot \nabla, & \nabla^\bot = \nabla - B_\mathrm{v}^{-1} \nabla \chi \partial^\parallel \\
\Delta^\bot = \nabla \cdot \nabla^\bot, & \Delta^* = B_\mathrm{v}^{-2}\nabla \cdot (B_\mathrm{v}^2 \nabla^\bot \\
\left[f, g\right] = B_\mathrm{v}^{-1} \nabla \chi \cdot \left(\nabla f \times \nabla g \right), & (f, g) = \nabla^\bot f \cdot \nabla^\bot g
\end{array}    
\end{eqnarray*}

These equations are similar to those used in \citep{ramasamy2024nonlinear}, except that a poloidal momentum source based on the background potential profile, $\Phi_0(s)$, has been introduced in the second line of equation \ref{eq:vorticity_equation}. In addition, the diffusion operators acting on density and temperature can be maintained by subtracting the initial equilibrium profiles $\rho_0(s)$ and $T_0(s)$ in equation \ref{eq:density_equation} and \ref{eq:temp_equation}. By default, the code does not maintain the background profiles, and these terms are not used. They have been applied in a subset of the simulations in Section \ref{sec:w7as}.

\section{Simulation parameters} \label{app:params}

\begin{table}
    \centering
    \caption{Physical and resolution parameters for baseline W7-X-like and W7-AS simulation in Section \ref{sec:w7x} and \ref{sec:w7as}, respectively. The approximate simulated values are computed at the magnetic axis of the initial equilibrium.}
    \begin{tabular}{c|c|c}
    Parameter & W7-X-like & W7-AS  \\ \hline
            $T\ [\mathrm{keV}]$                                                &     8.96               &    0.18                   \\
            $n\ [\times 10^{20}\mathrm{m}^{-3}]$                               &     0.341              &    2.0                    \\
            $\eta\ [\Omega \mathrm{m}]$                                        & $1.93 \times 10^{-8}$  &  $3.49 \times 10^{-7}$    \\
            $\eta_{\mathrm{num}}\ [\Omega \mathrm{m}^3]$                       & $1.93 \times 10^{-14}$ &  $3.49 \times 10^{-13}$   \\
            $\mu_\perp\ [\mathrm{kgm}^{-1}\mathrm{s}^{-1}]$                    & $1.55 \times 10^{-7}$  &  $6.46 \times 10^{-9}$    \\
            $\mu_{\mathrm{num}}\ [\mathrm{kgms}^{-1}]$                         & $1.55 \times 10^{-13}$ &  $6.46 \times 10^{-15}$   \\
            $\mu_{\parallel,\ \bot}\ [\mathrm{kgm}^{-1}\mathrm{s}^{-1}]$       & $2.87 \times 10^{-5}$  & $2.97 \times 10^{-7}$     \\
            $\mu_{\parallel,\ \parallel}\ [\mathrm{kgm}^{-1}\mathrm{s}^{-1}]$  & $2.87 \times 10^{-5}$  & $2.97 \times 10^{-7}$     \\
            $D_\bot [\mathrm{m}^2\mathrm{s}^{-1}]$                             & $0.154$                & $0.231$                   \\
            $\kappa_\bot [\mathrm{m}^2\mathrm{s}^{-1}]$                        & $0.231$                & $0.924$                   \\
            $\kappa_\parallel [\mathrm{m}^2\mathrm{s}^{-1}]$                   & $0.231\times 10^7$     & $0.231\times 10^7$        \\
            $S_e\ [\mathrm{MW}]$                                               & $0-20$                 &   $0-3.2$                 \\ 
            $\mathrm{n}_\mathrm{rad}$                                          & $64$                   &    $46$                   \\
            $\mathrm{n}_\mathrm{pol}$                                          & $64$                   &    $64$                   \\
            $\mathrm{n}_\mathrm{tor}$                                          & $0-30$                 &    $0-30$                
      \end{tabular}
      \label{tab:sim_params}
\end{table}

The simulation parameters for the studies in Section \ref{sec:w7x} and \ref{sec:w7as} are shown in Table \ref{tab:sim_params}. In both cases, the pressure profile is linearly decreasing towards the plasma boundary. The density profile is assumed constant in the W7-X-like case, and all of the viscoresistive and diffusive parameters are constant across the plasma volume. In the W7-AS case, the pressure is split between the temperature and density profiles assuming $n=n_0 \sqrt{\frac{p(s)}{p_0}}$, and $n_0$ refers to the value at the magnetic axis in Table \ref{tab:sim_params}. \textcolor{black}{With respect to the poloidal resolution parameters, it should be noted that JOREK uses cubic B\'ezier finite elements, such that the simulated broad mode structures in these low shear configurations are well resolved.}

\bibliographystyle{jpp}
\bibliography{references}

\end{document}